\documentstyle[11pt,aaspp4]{article}
\def\HI{\ion{H}{1}}

\def\sb{\ifmmode{\;{\rm mag}\;{\rm arcsec}^{-2}}\else{~mag~arcsec$^{-2}$}\fi}
\def\linsb{\ifmmode{\;L_{\sun}{\rm pc}^{-2}}\else{~$L_{\sun}{\rm pc}^{-2}$}\fi}
\def\csb{\ifmmode{\mu_0}\else{$\mu_0$}\fi}
\def\lincsb{\ifmmode{\Sigma_0}\else{$\Sigma_0$}\fi}
\def\solar{\ifmmode{_{\sun}}\else{$_{\sun}$}\fi}
\def\eg{e.g.,\ }
\def\ie{i.e.,\ }

\def\etal{{et al}.\ }
\def\MLo{\ifmmode{\Upsilon_o}\else{$\Upsilon_o$}\fi}
\def\ML*{\ifmmode{\Upsilon_*}\else{$\Upsilon_*$}\fi}
\def\MLT{\ifmmode{\Upsilon_T}\else{$\Upsilon_T$}\fi}
\def\kms{\ifmmode{\;{\rm km}\,{\rm s}^{-1}}\else{~${\rm km}\,{\rm s}^{-1}$}\fi}
\def\mass{\ifmmode{\cal M}\else{${\cal M}$}\fi}
\begin{document}

\title{Testing the Hypothesis of Modified Dynamics
       with Low Surface Brightness Galaxies and Other Evidence}
\author{Stacy S. McGaugh\altaffilmark{1}}
\affil{Department of Terrestrial Magnetism \\ Carnegie Institution of
Washington \\ 5241 Broad Branch Road, NW \\ Washington, DC 20015}
%\\ e-mail:  ssm@dtm.ciw.edu}
\and
\author{W. J. G. de Blok\altaffilmark{2}}
\affil{Kapteyn Astronomical Institute \\
Postbus 800 \\ 9700 AV Groningen \\ The Netherlands}
%\\e-mail: blok@astro.rug.nl}

\altaffiltext{1}{Present Address: Physics and Astronmy, Rutgers University,
136 Frelinghuysen Road, Piscataway, NJ 08854-8019}
\altaffiltext{2}{Present Address: Astrophysics Groups, School of Physics,
University of Melbourne, Parkville, Victoria 3052, Australia}

%\lefthead{McGaugh \& de Blok}
%\righthead{Testing MOND}

\begin{abstract}

The rotation curves of low surface brightness galaxies provide a unique
data set with which to test alternative theories of gravitation
over a large dynamic range in size, mass, surface density, and acceleration.
Many clearly fail, including any in which the mass discrepancy appears at a
particular length-scale.  One hypothesis, MOND [Milgrom 1983, ApJ, 270, 371],
is consistent with the data.  Indeed, it accurately predicts the observed
behavior.  We find no evidence on any scale which clearly contradicts MOND,
and a good deal which supports it.

\end{abstract}

\keywords{cosmology: dark matter --- galaxies: formation --- galaxies: halos ---
galaxies: kinematics and dynamics --- galaxies: structure --- gravitation}

\begin{quote}
\raggedleft
%\dots when
When you have eliminated the impossible, whatever remains, \\
however improbable, must be the truth. \\
--- Sherlock Holmes%, in {\it Sign of Four\/} by Sir Arthur Conan Doyle
\end{quote}

\clearpage

\section{Introduction}

There exists clear evidence for mass discrepancies in extragalactic
systems.  Application of the usual Newtonian dynamical equations to the
observed luminous mass does not predict the observed motions.
This leads to the inference of dynamically dominant amounts of dark matter.

In a preceding paper we described the many difficulties that arise
in trying to understand the data in terms of dark matter
(McGaugh \& de Blok\markcite{pI} 1998; hereafter paper I).
In this paper we consider the alternative of a change in the fundamental
equations of motion as the cause of the observed mass discrepancies.
Many alternative theories of gravity have been posited to this end,
but most have been ruled out (\eg Sanders 1986).
A few which may still appear to be viable are not in the light
of data for Low Surface Brightness (LSB) galaxies.
Length-scale dependent alteration of the inverse square law
(\eg the generic cases discussed by Liboff 1992 or the
nonsymmetric gravity of Moffat \& Sokolov 1996) can not explain the
large variation in the scale on which the mass discrepancy appears
(Fig.~3 of paper I)
unless the length-scale of the theory is allowed to vary from galaxy to
galaxy.  Similarly, the linear potential theory explored by Mannheimm \&
Kazanas (1989) predicts rotation curves which should ultimately rise rather
than remain flat.  This is not obviously consistent with extant rotation
curve data (Carlson \& Lowenstein 1996), and the rotation curves of at
least some LSB galaxies (\eg UGC 128) remain flat well beyond the point where
the upturn should be observed.  Only the Modified Newtonian
Dynamics (MOND) proposed by Milgrom (1983a,b,c) appears
empirically viable, so we focus on testing it.

The rotation curves of disk galaxies derived from gaseous tracers
(H$\alpha$ and \HI) provide the strongest tests of
alternative force-laws (\eg Kent\markcite{K87} 1987; Begeman\markcite{BBS}
\etal 1991).  With only the assumption of circular motion,
it is possible to directly equate the centripetal
acceleration
\begin{equation}
a_c = \frac{V^2}{R}
\end{equation}
with the gravitational acceleration
\begin{equation}
g = - \frac{\partial \varphi}{\partial R}
\end{equation}
determined from the Poisson equation
\begin{equation}
\nabla^2 \varphi = 4 \pi G \rho
\end{equation}
or any hypothesized alternative, for example
\begin{equation}
{\vec\nabla}\cdot\left[\mu\left(\frac{|\nabla\varphi|}{a_0}\right)
{\vec\nabla}\varphi\right] = 4 \pi G\rho
\end{equation}
(Bekenstein \& Milgrom 1984).
In no other type of system are tests so direct and free of assumptions.

To test MOND we employ the data compiled in paper I (mostly from
Broeils 1992, van der Hulst \etal 1993, and de Blok \etal 1996).
These are rotation curves of disk galaxies spanning a large range in size,
luminosity, and surface brightness.  Luminosity and surface brightness
must trace mass and mass surface density modulo only the mass-to-light
ratio of the stars in this alternative to dark matter.

Previous tests of MOND have given mixed results.  MOND does generally do
a good job of fitting rotation curves (Begeman \etal 1991), but it is
less clear that it works in systems other than disks (\eg clusters of
galaxies, The \& White 1988).  As emphasized by Milgrom (1983b,1988),
LSB galaxies provide particularly strong tests.  Some attempts to test
MOND with LSB dwarf galaxies found that it failed (\eg Lake 1989).
Milgrom (1991) pointed out some uncertainties and limitations
in Lake's analysis, so it remains unclear whether these cases
constitute contradictions of MOND.  Other attempts to fit a limited
number of LSB galaxies reported success (Milgrom \& Braun 1988;
Begeman \etal 1991; Sanders 1996).

We test MOND with the substantial amount of new data we have accumulated
on LSB galaxies (van der Hulst \etal 1993; de Blok \etal 1996).
Section 2 describes the predictions of
MOND relevant to LSB galaxies and tests each.  Section 3 examines the viability
of MOND in other systems.  A summary of the results and some discussion of
their implications is given in \S 4.  Detailed MOND fits to the rotation
curves of LSB galaxies are given by de Blok \& McGaugh 1998 (paper III).
Symbols and definitions follow the conventions of paper I.
We adopt $H_0 = 75\kms\,{\rm Mpc}^{-1}$ throughout.

\section{The Modified Dynamics}

The Modified Newtonian Dynamics are an empirically motivated force law
which can be interpreted either as a modification
of gravity or of the law of inertia (see review by Milgrom 1994).
We are not interested here in which interpretation is preferable,
but rather in testing whether MOND is indeed the correct force law.
In this context it should be realized that MOND has not yet been developed
into a full theory in the sense of General Relativity.  However, as a force
law it makes very precise and testable predictions just as the inverse
square law did well before the adornment of subsequent theoretical elaboration.

The empirical motivation for MOND is the observation that the rotation curves
of high surface brightness (HSB) spiral galaxies are asymptotically flat.
One must take care not to test only the fact the force law was constructed
to realize.  The properties of LSB galaxies were largely unknown at the
time, and were not part of the input motivating MOND.  Milgrom (1983b)
stated that ``{\it disk galaxies with low surface brightness
provide particularly strong tests}'' and made
a series of specific predictions about LSB galaxies which constitute
genuine tests of the MOND force law.

\subsection{The Modified Force Law}

The MOND force law is
\begin{equation}
a = \sqrt{g_N a_0},
\end{equation}
where $g_N$ is the usual Newtonian acceleration and $a_0$ is a universal
constant.  This applies only for $a \ll a_0$; for $a \gg a_0$ the behavior
is purely Newtonian.  There is an interpolation function $\mu(x)$ with
$x = a/a_0$ for connecting the two regimes (Milgrom 1983a).  The function
is required to have the asymptotic behavior $\mu(x \gg 1) \rightarrow 1$
and $\mu(x \ll 1) \rightarrow x$.  We will not specify $\mu(x)$
here, and restrict our tests to the deep MOND regime where
equation~5 is the effective force law.  In practice this means
$x < \onehalf$, where the precise form
of $\mu(x)$ ceases to matter (\ie $[\mu(x)-x]/x < 10\%$).
The MOND limit only occurs for extremely small accelerations:
$a_0 = 1.2 \times 10^{-10}\;{\rm m}\,{\rm s}^{-2}$ (Milgrom \& Braun 1988;
Begeman \etal 1991).
This is roughly 1 \AA ngstrom per second per second, or $10^{-11}\;
{\rm g}_{\oplus}$.

For circular orbits about a point mass, $\mass$, in the MOND limit,
\begin{equation}
\frac{V^2_c}{R} = \sqrt{\frac{G \mass}{R^2} a_0}.
\end{equation}
This gives an asymptotically constant rotation velocity $V_c$
independent of $R$:
\begin{equation}
V_c^4 = a_0 G \mass.
\end{equation}
It is this behavior that gives rise to asymptotically flat rotation curves
and the Tully-Fisher relation (Tully \& Fisher 1977).

Since MOND is a force law based on a modification at a particular
acceleration-scale, the strongest tests of MOND are provided by systems
with the lowest accelerations.
The stars in HSB galaxies experience centripetal accelerations of order $a_0$.
Low surface brightness galaxies have very low mean accelerations,
typically $\langle a \rangle \approx 10^{-11}\; {\rm m}\,{\rm s}^{-2} \sim
a_0/10$.  As a consequence, LSB galaxies provide strong tests of MOND.

\subsection{Testing MOND}

Solar system tests can be very precise, but do not probe the MOND regime.
The transition between Newtonian and MOND regimes allows us to define a
transition radius where $a = a_0$,
\begin{equation}
R_t = \sqrt{G \mass/a_0}.
\end{equation}
For the sun, $R_t \approx 0.1$ light year.  One would not expect to notice
MOND effects until this radius, at which point
the field of the Milky Way becomes significant.
Classic solar system tests near to the sun are
restricted to the regime $a \gg a_0$ and only test the deviation
of the interpolation function from its asymptotic limit (Milgrom 1983a).
The same holds true for binary pulsars, which experience large
accelerations well removed from the MOND regime.

Extragalactic systems are the only laboratories where the MOND regime
is clearly probed.  In fitting MOND to galaxy data, there are several
adjustable parameters.  One is the acceleration constant itself.
This must be a universal constant; once measured accurately in one galaxy
the same value must apply in all galaxies.  We adopt $a_0 = 1.2\; {\rm \AA}\,
{\rm s}^{-2}$ and keep it fixed.  This was determined by Begeman \etal (1991)
to be the best fit to a sample of galaxies with high quality \HI\ rotation
curves which has no overlap with our new LSB galaxy data.  Slight
adjustments to the distance to a galaxy can sometimes improve a MOND fit
(Begeman \etal 1991).  This occurs because of the interplay between the
MOND mass, which depends only on the distance-independent circular velocity,
and the gas mass, which varies as $D^2$.
We keep $D$ fixed at the value determined from the assumed $H_0$.  For
other choices of $H_0$ there is some limited freedom to adjust $a_0$ to
compensate.  As with $a_0$ and $D$, adjusting the inclination of a disk galaxy
can improve some fits.  Inclination is crucial since MOND
masses depend on $V^4/\sin^4(i)$.  In this paper we keep $i$ fixed; see
paper III for further examination of this point.

So far we have named three parameters which could in principle be adjusted.
In practice, there is very little freedom to do this, and
we will keep all of them fixed.  The only truly free parameter is the
conversion factor from light to mass, \ML*.  We will treat this as a
value to be determined and compared to the expectations for stellar
populations.

In the following subsections, we discuss particular tests of MOND with
LSB galaxies.

\subsection{The Tully-Fisher Relation}

A very strong prediction of MOND is a single universal Tully-Fisher relation.
``{\it The relation between asymptotic velocity and the mass of the galaxy
is an absolute one.}'' (Milgrom 1983b).
This follows from equation~(7) which gives
\begin{equation}
V_c^4 = a_0 G \mass = a_0 G \left(\frac{\mass_b}{L}\right) L
\propto \Upsilon_b L
\end{equation}
where the baryonic mass-to-light ratio $\Upsilon_b = \mass_b/L$.
Note that $R$ does not appear in this equation, nor do the dark matter
galaxy formation parameters of equation~(13) in paper I.

That $R$ does not appear in equation~(9) is of fundamental significance.
In the case of purely Newtonian dynamics, $V^2 = G \mass/R$.  Since LSB
galaxies have, by definition, larger $R$ than HSB galaxies of the same
luminosity, they should not fall on the same Tully-Fisher relation.
Yet they do (Sprayberry \etal 1995; Zwaan \etal 1995).  Invoking a
length-scale for the mass which is different from that of the light
helps not at all: serious fine-tuning problems occur in the
dark matter picture (paper I).  In MOND the Tully-Fisher relation
follows simply and  directly from the form of the force law.

The error budget for scatter in the Tully-Fisher relation in magnitudes is
\begin{equation}
\left|\delta M\right|^2 = 18.86 \left|\frac{\delta V_c}{V_c}\right|^2
+ 1.18 \left|\frac{\delta \Upsilon_b}{\Upsilon_b}\right|^2.
\end{equation}
Comparison of this with equation~12 of paper I shows that the term involving
the central surface brightness has disappeared.  The real difference is larger,
though.  In the dark matter picture, there should be some intrinsic variation
in $V_c$ from halo to halo.  In MOND, mass and velocity are strictly related;
the term for the error in $V_c$ in equation~(10) now refers only to
observational uncertainties and not also to intrinsic variations.
All but one of the terms which should contribute intrinsic variance in
the dark matter picture disappear.  To obtain a single universal Tully-Fisher
relation with little scatter all that is really required in MOND
is that $\mass_b/L$ be roughly constant:
$\delta M = 1.086\, |\delta \Upsilon_b/\Upsilon_b|$.
Small intrinsic scatter in the Tully-Fisher relation is much easier to
understand with MOND than with dark matter.

The slope of the Tully-Fisher relation, $L \propto V^y$ with $y=4$
(Aaronson \etal 1979; Tully \& Verheijen 1997; paper I),
is dictated by the form of the force law in MOND.
No particular slope is required in the case of dark matter.
A slope of $y = 4$ is sometimes attributed to the virial theorem
(\eg Silk 1997), but a slope closer to $y = 3$ has also been suggested as the
value arising in plausible disk galaxy formation scenarios
(\eg Mo \etal 1997).  In any case, a constant disk plus halo mass surface
density must be arranged.  Multi-component disk-halo models generally
predict that some signature of the disk component should be visible in
Tully-Fisher residuals, but none are present (Courteau \& Rix 1997).  More
generally, there is no signature of the transition from disk to halo domination
(the ``disk-halo conspiracy'').  These all occur naturally if there is no
halo and the disk is the only component, as in MOND.

The Tully-Fisher relation of LSB galaxies provides a new test of MOND.
That HSB galaxies obey the Tully-Fisher relation was known when MOND was
developed, so this does not itself constitute a test.  That LSB
galaxies fall on the same relation with the same normalization was not known.
That they should constitutes a genuine prediction (Milgrom 1983b):
``{\it We predict, for example, that the proportionality factor in
the $\mass \propto V_c^4$ relation for these galaxies is the same as
for high surface density galaxies.}''

\subsection{The $\Upsilon$-$\Sigma$ Conspiracy}

If the MOND force law is the basis of the Tully-Fisher relation,
it should be possible to derive from MOND the $\Upsilon$-$\Sigma$
conspiracy inferred with conventional dynamics (Zwaan \etal 1995; paper I).
This requires $\MLo^2 \Sigma \propto$ constant, where \MLo\ is the
conventional dynamical mass-to-light ratio.
In MOND, the typical acceleration of a disk is proportional
to the square root of its characteristic surface density so that
\begin{equation}
\langle a \rangle^2 \propto \sigma = \Upsilon_b \Sigma
\end{equation}
(equation 6 of Milgrom 1983b and equation~19).

The severity of the inferred Newtonian mass discrepancy
$\mass_N/\mass_b$ depends on the extent to which the acceleration
is in the MOND regime:
\begin{equation}
\frac{\mass_N}{\mass_{b}} = \frac{1}{\mu(x)} = \frac{\MLo}{\Upsilon_{b}}
\approx \frac{a_0}{\langle a \rangle}.
\end{equation}
The approximation effectively becomes an equality
for $\langle a \rangle < a_0/2$.  As a result,
$\MLo = \Upsilon_b a_0/\langle a \rangle$.  Using equation~(11),
this becomes $\MLo^2 \propto (a_0^2/\sigma)\Upsilon_b^2 =
a_0^2 \Upsilon_b/\Sigma$, \ie
\begin{equation}
\MLo^2 \Sigma \propto \Upsilon_b a_0^2 \sim {\rm constant}.
\end{equation}
The $\Upsilon$-$\Sigma$ conspiracy does follow from MOND.

\subsection{Stellar Mass-to-Light Ratios}

Another test of MOND is provided by the values required for the
mass-to-light ratio of stellar populations, \ML*.  These should be consistent
with what we know about stars.  Since there is no dark matter, the
kinematic data provide a direct measure of the luminous mass.

The only free parameter in equation~(9) is \ML*.  The constants $a_0$ and
$G$ are known, $V_c$, $L$, and the gas mass $\mass_g$ are measured.  Hence,
$\mass_* = \mass - \mass_g$ and $\ML* = \mass_*/L$ where $\mass$ is the
total dynamical mass indicated by MOND.
For spiral disks we expect \ML*\ in the range of 1 -- 2 in the $B$-band,
with the range $0.5 < \ML* < 4$ being credible (\eg Larson \& Tinsley 1978;
Bruzual \& Charlot 1993).

The gas mass is inferred directly from the 21 cm flux multiplied by
a conversion factor to account for helium and metals:
$\mass_g = 1.4 \mass_{HI}$.  The conversion factor is probably closer
to the primordial value 1.32 for LSB galaxies since these systems
are very metal poor (McGaugh 1994; R\"onnback \& Bergvall 1995).
This difference is very small compared to the accuracy with which \ML*\ can
be estimated.  We assume molecular gas is not a major mass component.
High surface brightness galaxies are gas poor with molecular
gas mass being a small fraction of the stellar mass (Young \& Knezek 1989;
McGaugh \& de Blok 1997).
Low surface brightness galaxies are \HI\ rich but apparently have very
little molecular gas (Schombert \etal 1990; de Blok 1997).
Any molecular mass which is
present will be attributed to stars, resulting in \ML*\ which are slightly
too high.  Again, this is a small effect compared to the uncertainty in \ML*.

Strictly speaking, equation~(7) holds precisely only for point masses.
Disk galaxies are reasonably well approximated by an exponential mass
distribution for which
\begin{equation}
\mass(R) = 2 \pi \sigma_0 h^2 \left[1-e^{-R/h}
\left(1+\frac{R}{h}\right)\right]
\end{equation}
and
\begin{equation}
g_N(R) = \pi G \sigma_0 \frac{R}{h} \left[I_0(R/2h) K_0(R/2h) -
I_1(R/2h) K_1(R/2h)\right],
\end{equation}
where $I_n$ and $K_n$ are modified Bessel functions of the first and
second kind (Freeman 1970).
Combining these and following the MOND prescription $a= \sqrt{g_N a_0}$
gives a dimensionless factor $\chi(R)$
correcting the simple point mass formula for the disk geometry:
\begin{equation}
\chi(R) = \frac{2 h^3 \left[1-e^{-R/h} (1+R/h)\right]}
{R^3 \left[I_0(R/2h) K_0(R/2h)
- I_1(R/2h) K_1(R/2h)\right]}
\end{equation}
so that
\begin{equation}
\mass(R) = \chi(R) \frac{V^4(R)}{a_0 G}.
\end{equation}
This is a fairly mild correction, as $\chi(R)$ does not deviate far from
unity at most radii.  For consistency with paper I we evaluate it at
$R= 4h$, where $\chi(4h) = 0.76$.  The effect is
to reduce the mass somewhat as a disk geometry rotates faster
than the equivalent spherical mass distribution.

\placetable{MONDML}

The stellar mass is computed as
\begin{equation}
\mass_* = 0.76 \frac{V_c^4}{a_0 G} - 1.4 \mass_{HI}.
\end{equation}
We have determined \ML*\ in this fashion for all galaxies meeting
the requirements stipulated in paper I (essentially everything for
which we have a measurement of $V_c$, $L$, \csb, and $h$).
The results are given in Table~1 and plotted in Fig.~1, which is analogous
to Fig.~4 of paper I.

\placefigure{MLmond}

The stellar mass-to-light ratios inferred from MOND are consistent
with those expected for disk population stars.
In the mean, $\langle \ML* \rangle = 1.9$,
and $\ML* = 1.6$ in the median.  There is a fair amount of scatter about
the mean value.  No trend of \ML*\ with \csb\ is obvious in Fig.~1(b).
Slight trends of \ML*\ with $M_B$ and $h$ are perceptible in Figs.~1(a) and
(c), but are not highly significant as they are strongly influenced by the low
\ML*\ values of a few small galaxies.  The directions of these trends are
nevertheless consistent with those of color-magnitude and color-size relations.

MOND can be quite sensitive to the parameters we have held fixed:
$a_0$, $D$, and $i$.  It has not been necessary to adjust any of these
parameters to get reasonable results from the global approach taken here.
By global, we mean that we use only the asymptotic flat circular velocity
$V_c$ to determine \mass.  In paper III we present
fits to the full shape of $V(R)$.  This is a different exercise which gives
slightly different results for the mass-to-light ratios of individual galaxies.

The precise distance to a galaxy is sometimes important because of the
different $D$-dependence of gas mass and total MOND mass.
Total mass $\mass \propto V_c^4$ is measured independent of distance (for
small redshifts), but $\mass_g \propto D^2$.  An overestimate of the
distance can lead to a situation in which it appears that the gas mass
exceeds the total dynamical mass.
For a sample with many gas rich galaxies, it would be surprising if
this did not occasionally manifest itself.  Indeed, it is well known in
the case of DDO 154 (Milgrom \& Braun 1988; Begeman \etal 1991;
Sanders 1996).  Examination of Table~1 shows that F571--V1 and F574--1
might be similar cases, as both are inferred to have unreasonably
small or negative \ML*.  A 20\% reduction in the assumed distance to
these galaxies would have the effect $\ML*($F571--V1$) \rightarrow 0.5$
and $\ML*($F574--1$) \rightarrow 0.8$.  Given the uncertainty in the
distance scale, neither case provides a clear counterexample to MOND.

MOND is also very sensitive to the inclination since the observed line of
sight velocity must be corrected by $\sin^{-4}(i)$.
Inclination determinations can be difficult and are sometimes quite
uncertain for LSB galaxies (McGaugh \& Bothun 1994; de Blok \etal 1995;
de Blok \etal 1996).  In the case of F571--V1, \ML*\ becomes
positive if $i = 35 \rightarrow 34$.  We certainly would not claim to be
able to determine inclinations this accurately by information independent
of MOND (paper III).

In the case of F574--1, it is not obvious that the
asymptotically flat part of the rotation curve has been reached (de Blok \etal
1996).  If $V_c$ is underestimated, the MOND mass will be too small.
The stellar mass to light ratio becomes more reasonable ($\ML* > 0.5$)
if $V_c = 91 \rightarrow 99\kms$, which is possible within the errors.
It is also possible that the factor $\chi(R)$ we have assumed could be
an underestimate in this case.

In sum, MOND yields reasonable values for the mass-to-light ratios
of stellar populations.  This is not a trivial result.  Since $\ML* \propto
V_c^4$, even small errors in $V_c$ can make \ML*\ seem incorrect.
There are a few individual galaxies for which \ML*\ may seem a bit
unreasonable, but is this surprising given the nature of astronomical data?

\subsection{Mass Surface Densities}

Since there is no dark matter in the MOND hypothesis, mass surface
density must be correlated with luminous surface density.
A way to see this without invoking stellar populations
is through the characteristic parameter $\xi$
defined by Milgrom (1983b) for exponential disks:
\begin{equation}
\xi \equiv \frac{V_c^2}{a_0 h} = \sqrt{\frac{G \mass}{a_0 h^2}}.
\end{equation}
This is proportional to the characteristic acceleration $V_c^2/h$ of a disk
and to the square root of its mass surface density ($\sigma \propto \mass/h^2$).

\placefigure{Ximond}

By using the observed velocity and scale length, we can construct the
quantity $\xi$ which provides a direct dynamical estimate of the mass
surface density.  Values computed for $\xi$ are given in Table~1.
The mass surface density indicated dynamically by $\xi$
must be correlated with with the mass surface density of ordinary matter,
indicated by the central surface brightness \csb.
These quantities are plotted against each other in Fig.~2.
There is a strong correlation, with regression coefficient ${\cal R} = 0.82$.
The slope and normalization of the relation is consistent with that expected
in MOND.  From equation~(19), one can derive
\begin{equation}
\log(\xi) = -\frac{1}{5}(\csb - 27) +
\frac{1}{2} \log\left(\frac{2\pi G}{a_0} \Upsilon_b\right)
\end{equation}
assuming an exponential mass distribution with $\mass = 2\pi \sigma_0 h^2$.
A formal fit to the data gives a slope of $-0.21$, consistent with the
expected $-0.2$.  The normalization is also correct for
$\Upsilon_b \approx 4$, a reasonable value given the
assumption of an exponential mass distribution.  Deviations from this
idealized case will alter the factor $2 \pi$ in equation~(20) and hence
the precise value of $\Upsilon_b$ that is inferred.

\subsection{Conventional Quantities}

The effects of MOND, when interpreted in terms of conventional dynamics,
lead to specific predictions for the behavior of conventional quantities.
For example, the Newtonian mass-to-light ratio in any given galaxy
should show no mass discrepancy until beyond the transition
radius defined by equation~(8).  After this point, $\MLo(R)$ should increase
as $R$ increases.  As one examines galaxies of decreasing
central surface brightness, two effects should be apparent:  the
severity of the discrepancy should grow, and the radius (measured in
scale lengths) at which the discrepancy sets in should shrink.  If at
every point in a galaxy $a(R) \ll a_0$, the mass discrepancy should be
apparent at essentially $R = 0$ (Milgrom 1983b).

\placefigure{MLR}

In Fig.~3 we plot \MLo\ against $R/h$ for all available galaxies
(Tables 1 and 2 of paper I).  The mass-to-light ratio is computed as
the accumulated mass within radius $R$ divided by the accumulated
luminosity at the same point.  Mass is assumed to be distributed
spherically [$\mass \propto R V^2(R)$] which is an adequate approximation
for present purposes if not strictly true at small radii.
We have subdivided the data in Fig.~3 into
one-magnitude bins in surface brightness.
The rise of \MLo\ does appear to become more pronounced as \csb\ declines.
There is no clearly defined transition radius, but all disk galaxies
have $a \lesssim a_0$.  This may be a requirement for disk stability
(see \S 3.3).

\placefigure{AR}

What really matters is the acceleration.  Another way of stating the
prediction of MOND is that $\langle a \rangle$ should decrease with surface
density (equation~19).  In Fig.~4 we plot the rotation curves as the centripetal
acceleration required to produce them: $a = V^2(R)/R$.  There is some scatter
induced by inclination uncertainties and intrinsic scatter in \ML*.
Nevertheless, there is a clear trend for $\langle a \rangle$ to decline
with surface brightness, as expected in MOND.

\subsection{Rotation Curve Shapes}

The shapes of the rotation curves of exponential disks is controlled in MOND
by the parameter $\xi$
which is closely related to the surface brightness.  Milgrom (1983b) predicts
``{\it a correlation between the value of the average surface density
(or brightness) of a galaxy and the steepness with which the rotational
velocity rises to its asymptotic value.  Small surface densities imply slow
rise of V.}''
We noted in paper I the slow rate of rise of the rotation curves of
LSB galaxies (fact 2).  Our empirical statement is essentially identical
to the prediction of Milgrom (1983b).  As a measure of this effect,
Milgrom (1983b) suggested the radius (in scale lengths) at which $V(R/h) =
V_c/2$.  This contains the same information as $R_{34}$ in paper I which
should thus be correlated with surface brightness.  That is, Fig.~12(b) of
paper I is in fact a prediction of MOND.

We illustrate this further by plotting the rotation curves logarithmically to
show their shape (Fig.~5), binned by surface brightness as before.
The rotation curves
of the highest surface brightness galaxies rise rapidly, often reaching
the asymptotic velocity within 1 scale length.  The rate of rise becomes
slower as surface brightness decreases.  The asymptotic flat velocity
is only reached at 2 or more scale lengths in the lowest surface brightness
galaxies, as expected in MOND.

\placefigure{shape}

Another systematic of rotation curves is that the rate of rise of $V(R/h)$
is well correlated with absolute magnitude
(Fig~12a of paper I).  That this occurs is not predicted or required by MOND.
Hence it neither supports nor contradicts MOND.

Though the rotation curve shapes are consistent with MOND, the predictions
discussed in this section are less strong than previous ones.  The systematic
trend of $R_{34}/h$ with $\xi$ is predicted to be fairly weak, and the
presence of a bulge component can have an additional effect on the shape
of $V(R)$ not accounted for by the quantity $\xi$ defined for exponential
disks.  Though bulges are generally anemic in LSB galaxies, they are present
in the higher surface brightness galaxies in Fig.~5.
The question then becomes whether MOND can
fit the rotation curves in detail given the actual observed luminous
mass distribution and not just the exponential disk approximation.

\subsection{Residuals of Rotation Curve Fits}

MOND is known to work well in fitting rotation curves, particularly in HSB
galaxies (Begeman \etal 1991; Sanders 1996).  Except for a few dwarfs in
those samples, $V(R)$ reaches the asymptotic flat velocity very rapidly.
This is not true in LSB galaxies:  MOND must not only give an asymptotically
flat rotation curve, but it must also give the gradual observed rise.
It was not contrived to do this, so detailed fits to LSB galaxy rotation curves
provide strong tests.

\placefigure{f5631}

An example of a detailed MOND fit to an LSB galaxy is given in Fig.~6.
Fits for the entire sample are the subject of paper III.
We make use of those results here with only a few comments.
The prediction of $V(R)$ follows with one fittable parameter, \ML*.
The freedom in adjusting \ML*\ is very limited.  It does not have a strong
effect on the fit ($V \propto \ML*^{1/4}$) and it must return a
reasonable value for a stellar population.
The shape of the stellar mass distribution is constrained to be the same
as that of the light; \ML*\ acts only as a normalization factor
and is not allowed to vary with radius.
The gas mass is comparable to the stellar mass in these
galaxies.  There is very little uncertainty in the conversion factor from
21 cm flux to \HI\ mass, and we have no freedom to adjust it.  The importance
of the gas component often constrains \ML*\ so tightly that in many cases
the ``fits'' are effectively parameter-free.  For nine LSB galaxies, good fits
were found immediately with one parameter (\ML* only) fits.  For six
others, achieving tolerable
fits required adjustment of the inclination (paper III).  MOND is
very sensitive to this since it enters through $\sin^{-4}(i)$ and a number
of the LSB galaxies in our sample are fairly face-on.
Two parameter (\ML*, $i$) MOND fits remain tightly constrained, and both
parameters are subject to independent checks.  In contrast,
dark matter fits require at least three parameters (\ML*\ and two halo
parameters).  The two halo parameters are not constrained by any data
independent of the rotation curves, leading to notorious degeneracies
(Athanassoula \etal 1987; Kent 1987; de Blok \& McGaugh 1997).

The residuals to the MOND fits to the rotation curves of LSB galaxies
from paper III are shown in Fig.~7 as a
function of the critical parameter $x = a/a_0$.  The data extend over
an order of magnitude range, all with $x < 1$ and most with $x < \onehalf$.
(See Sanders 1996 for fits to galaxies with $x > 1$ as well as $x < 1$).
Each point is a measured point from a rotation curve;
the residuals of all LSB galaxy fits are plotted together.
The data closely follow the line of zero residual with $\sim 10\%$
scatter, about what is expected for our errors.  In absolute terms,
the mean deviation $\Delta V = V_{obs} - V_{MOND} = 0.6 \pm 5.6\;{\rm km}\,
{\rm s}^{-1}$.

\placefigure{aca0}

The strongest test is provided by the points with $x < \onehalf$ where the
form of the interpolation function $\mu(x)$ is insignificant.
In this part of the diagram, not only is the mean residual small, but
there is no systematic trend of the residuals with $x$.  For $x > \onehalf$,
there is a hint of a systematic deviation.  The difference is not
highly significant, but it might indicate
that the actual from of the interpolation function is
slightly different than the one assumed.
The explanation may be more prosaic, as some of the
data are modestly affected by beam smearing (see de Blok \& McGaugh
1997 for an extensive discussion).

There is one further consequence of Fig.~7.
Even though the mass discrepancy does not appear at a particular
length scale, it does appear at a particular acceleration scale.
MOND is the {\it effective\/} force law in disk galaxies.

\section{Other Evidence}

The data for LSB galaxies are consistent with MOND.  Indeed, each of the
specific predictions Milgrom (1983b) made about LSB galaxies is confirmed.
This means something.  Is it possible that MOND, not dark matter, is the
solution to the mass discrepancy problem?

There are many other systems besides disks in which the mass discrepancy
is apparent.  If MOND is correct,
it must also be consistent with tests from these other systems.  Tests
are provided by any system with $\langle a \rangle \ll a_0$; in systems
where $\langle a \rangle \gg a_0$, no mass discrepancy should be inferred
unless there is genuine hidden mass.
Here we review the evidence provided by other types of systems on
different scales.

\subsection{Dwarf Spheroidals}

A class of galaxies which are low in surface brightness but
quite distinct from the LSB galaxies we have so far discussed are the
dwarf Spheroidals (dSph) found in the Local Group.  They are distinct
for a number of reasons, most importantly that they are three dimensional
systems supported by the velocity dispersion of their constituent stars,
not rotating disks.  As LSB systems, they should have low accelerations
and provide a strong test of MOND:
``{\it Effects of the modified dynamics are predicted to be particularly
strong in dwarf elliptical galaxies}'' (Milgrom 1983b).

That dSph galaxies have large mass discrepancies is clear
(Mateo \etal 1993; Vogt \etal 1995; Mateo 1996).
However, it is less clear that the discrepancies are of the correct
magnitude for consistency with MOND.  Gerhard \& Spergel (1992) and
Gerhard (1994) have applied this test.  In some cases, MOND seems to
do well, giving stellar mass-to-light ratios which are reasonable
for the stellar populations of these galaxies.  In other cases, the
results are less satisfactory.

Dwarf Spheroidals provide a strong test in the sense that the acceleration
is low, comparable to LSB galaxies.  However,
they do not provide as clean a test.  What one really needs to know
is the three dimensional velocity ellipsoid.  All that is possible,
of course, is to obtain a line-of-sight velocity dispersion $\varsigma$.
Additional assumptions are therefore required to interpret the observations.
We must assume that the MOND equivalent of the virial theorem applies
(Milgrom 1984; Gerhard \& Spergel 1992; Milgrom 1995) and
that each system is either spherical or isotropic.

In general, the mass of an isolated system in the MOND regime is
\begin{equation}
\mass = \frac{9}{4} \frac{\langle V_{rms}^2 \rangle^2}{a_0 G}
\end{equation}
(Milgrom 1995).  The assumption of sphericity or isotropy is necessary to
relate the {\it rms\/} velocity to the observable line-of-sight velocity
$\langle V_{rms}^2 \rangle = 3 \varsigma^2$.  Thus the effective equation is
\begin{equation}
\mass = \frac{81}{4} \frac{\varsigma^4}{a_0 G}.
\end{equation}
Note that the assumed relation between {\it rms\/} and line-of-sight velocity
has a significant impact in the geometric constant relating mass to velocity.
Since $\mass \propto \varsigma^4$,
a 20\% error in $\varsigma$ will lead to a factor of two error in mass.
This need not arise as an error.  It might
occur even with perfect observations simply because the true
velocity ellipsoid may not have the form we are obliged to assume.

Gerhard \& Spergel (1992) and Gerhard (1994) concluded that MOND fails
because it yields unreasonable \ML*\ for some dwarf Spheroidals, especially
Fornax and Ursa Minor.  By reasonable, we mean $1 < \ML* < 6$ in the $V$-band.
There is considerable uncertainty in the
stellar content of these galaxies.  Originally thought to be old implying
high \ML*, at least some appear surprisingly young (\eg Mateo \etal 1991;
Smecker-Hane \etal 1994).  The \ML*\ of Gerhard (1994)
are plotted in Fig.~8.  One can immediately see two things:
there are no error estimates, and many of the values
are in fact quite reasonable.  So, do the unreasonable values
falsify MOND, or are the errors simply very large in those cases?

\placefigure{dsph}

Data of the sort required for this analysis have accumulated rapidly
in the period during which these analyses were performed.  As a result,
rather better data now exist than did at the time of these analyses.
Milgrom (1995) has revisited this issue.  The results he obtained using
the same method and equations as Gerhard \& Spergel (1992) are also
plotted in Fig.~8.  Within the errors, the results give reasonable
values for \ML*.  The only thing that has really changed is that the data have
improved --- data which come from the same sources.

The data continue to improve, so we repeat this analysis using the
compilation of data provided by Mateo (1996).  Equation~(22)
above only applies for isolated spheres.  When a system is subject to
a dominant external field (\eg that of the Milky Way), the MOND behavior
is quasi-Newtonian
\begin{equation}
\mass = 2 G_{eff}^{-1} \, \varsigma^2 R_V
\end{equation}
with a modified effective constant of gravitation,
$G_{eff} \rightarrow a_0 G/\langle a \rangle$.
Here, $R_V$ is the virial radius (equation~10
of Gerhard \& Spergel 1992).  Which equation applies depends on the relative
accelerations imposed by the internal and external fields:
\begin{equation}
\eta = \frac{3\varsigma^2/2R_C}{V_{MW}^2/D}
\end{equation}
where $R_C$ is the core radius of the King profile which fits the light
distribution of the dSph in question, $D$ is its galactocentric distance,
and $V_{MW}$ is the asymptotic velocity of the Milky Way.  For the latter
we adopt $V_{MW} = 220\kms$.  If $\eta > 1$ the internal field dominates
and the object can be treated as
isolated (equation~22).  If $\eta < 1$, the external field due to the
Milky Way is dominant, and the behavior is quasi-Newtonian (equation~23).
As can be seen in Table~2, most dSph galaxies for which good kinematic data
exist are probably influenced by the external field of the Milky Way.
When $\eta \approx 1$, neither of these limits are really appropriate.
This is the case for many objects when the uncertainties in the many
parameters determining $\eta$ are considered.  A significant external field
can elongate the internal field and hence affect the assumed geometry,
so there are significant uncertainties besetting this analysis
regardless of the precision of the data.  Only Leo I provides a test in
a reasonably isolated system.

Where Mateo (1996) gives different values
for $\varsigma$, we adopt the one based on the greater number of stars.
In most cases, $\varsigma$ has only been measured in the central regions.
A global mean is more appropriate, so in the case of
Fornax where a velocity dispersion profile is available we use it to estimate
$\varsigma \approx 12\kms$.
We do not compute $\ML*$ for two dwarfs for which velocity dispersions
are listed, LGS 3 and Sagittarius.  The velocity dispersion
listed for LGS 3 gives $\ML* = 2.5$ but is based on only 4 stars, so the
uncertainties render this meaningless.  In the case of Sagittarius, the very
close interaction with the Milky Way undermines all of the assumptions on
which the analysis is based.

\placetable{dwarf}

Results of our analysis of the data compiled by Mateo (1996) are listed
in Table~2 and shown in Fig.~8 together with the previous determinations
of Gerhard (1994) and of Milgrom (1996).
Our error bars are based only on the error in $\varsigma$ given by Mateo (1996);
they do not include the large uncertainties in the luminosity, surface
brightness, and distance to these galaxies.  Neither do we attempt to
estimate any of the systematic uncertainties discussed above.  Hence the
error bars plotted in Fig.~8 are a lower limit to the true range of
uncertainty.  Nevertheless, they do show that our determinations are
consistent with previous ones, when the errors are considered.

Most dwarf Spheroidals do in fact lie in the range $1 < \ML* < 6$
according to all three independent analyses.  Marginal cases are
Draco and Ursa Minor (\ML*\ too high) and Fornax (\ML*\ too low).
The deviation of Draco and Fornax from the reasonable range is marginal
($\sim 1 \sigma$).  Draco has $\eta$ very near to unity which will tend
to cause a modest overestimate of \ML*\ (Milgrom 1995).  Ursa Minor is
the most problematic case, but even there the deviation is only
$\sim 2 \sigma$.  There is a complication in this case in that some
of the velocity dispersion may result from rotation (Mateo 1996).  This
can affect the analysis since rotational velocity enters the calculation
with less weight than does velocity dispersion.  Also, Mateo (1996)
lists two somewhat different velocity dispersions for this object,
both based on a sizable number of stars.  Armandroff \etal (1995) give
$\varsigma = 8.8 \pm 0.8\kms$ while Hargreaves \etal (1994) give
$\varsigma = 6.7 \pm 1\kms$.  Using the
latter value gives \ML*(Ursa Minor) $= 10^{+6}_{-4}$, not significantly
unreasonable.

Another complication is the systematic uncertainty in the parameters of
the Milky Way.  Since $\eta$ is close to unity in most cases, a difference
in the assumed strength of the external field of the Milky Way might make
a difference.  It has been suggested that $V_{MW}$ could be as low as
$180\kms$ (Olling 1997, private communication).  If we re-do the analysis
with this value, little changes.  In most cases there is a slight shift
towards lower \ML*.  The exception is Ursa Minor, where it makes a substantial
difference.  For the higher velocity dispersion of Armandroff \etal (1995),
$\ML* = 17 \rightarrow 10$.  In principle, uncertainties in the distance
to individual dwarfs can have similar effects.

Gerhard (1994) stresses that \ML*\ varies by a factor of 20 between
Fornax and Ursa Minor.  It is true that this is a lot, but it is also true
that these are the two most extreme points on
either side of a sensible mean value.  This is bound to
happen at some level, and the degree to which it occurs here
does not seem outrageous given the
uncertainty in the data and the analysis.

In sum, there is no evidence which clearly contradicts MOND in the data
for dSph galaxies.
Agreement between these data and MOND has generally improved as the
data have improved.  The qualitative prediction of MOND, that dSph
galaxies should have high mass discrepancies, is not
in dispute.  MOND also appears to work quantitatively.

\subsection{Giant Elliptical Galaxies}

The kinematic data which exist for giant elliptical galaxies is rather
limited.  This is because observations of these systems are generally
restricted to absorption lines which can typically be measured only
within one half light radius ($R_e$) where the surface brightness is
high (\eg Bertin \etal 1994).  Application of these data as a test of
any hypothesized force law is limited by the limited range of radii
probed.  It is further hindered by the fact that ellipticals are
quite complicated systems.  The brightest ellipticals are predominantly
pressure supported, but often have significant rotation as well
(Davies \etal 1983).  Even in the absence of rotation,
anisotropies in the velocity ellipsoid must exist to give rise to
the observed range of shapes of ellipticals.  This is quite
problematic for applying equation~(22): we simply do not have adequate
knowledge of the geometry of the orbits. 

For testing MOND, an even more severe constraint applies.  The typical
acceleration at radii for which measurements are available
(Bertin \etal 1994) is
\begin{equation}
a \approx \frac{3}{2} \frac{\varsigma^2}{R_e} \gtrsim 3 a_0.
\end{equation}
Extant data do not probe the MOND regime.

Giant ellipticals therefore provide no direct test of MOND.
They do provide several indirect tests, however.  One is that there
should be no apparent mass discrepancy where $a > a_0$.  This
is consistent with the persistent lack of
a clear need for dark matter in elliptical galaxies (van der Marel 1991;
Bertin \etal 1994).

Though the data may not probe the MOND regime, MOND effects must matter
at some level in elliptical galaxies.  One might expect some critical
phenomenon associated with the MOND scale $a_0$.  In the simple case
of spherical galaxies in the MOND limit, Milgrom (1983b) expects a
Faber-Jackson (Faber \& Jackson 1976) relation of the form
$\mass \propto \varsigma^4$.  If ellipticals are approximately isothermal,
then Milgrom (1984) also expects a Fish (1964) law.
MOND may therefore manifest itself in the regularity of the
Fundamental Plane (Djorgovski \& Davis 1987).  The analog of the
Fundamental Plane for disks is the Tully-Fisher relation.  With
\csb\ or $R_e$ as a third parameter, the ``fundamental plane'' of disks
is perpendicular to the luminosity-velocity plane, resulting in a narrow
Tully-Fisher relation for galaxies of all surface brightnesses.  Similarly,
the Fundamental Plane of ellipticals is viewed nearly edge-on in the
luminosity-velocity dispersion plane.  This suggests that the same
effect is at work.  That there is a modest tilt to the Fundamental
Plane merely indicates some modest systematic trend
of \ML*\ or the degree of pressure support with luminosity.

A generalization of the Fundamental Plane by Burstein \etal (1997) finds
that essentially all extragalactic systems form narrow structures in
``$\kappa$-space.'' This is indicative of a universal scale like $a_0$.
The $\kappa$-structures exhibit little surface density dependence,
the signature of the MOND force law.

Elliptical galaxies with shell systems provide an interesting probe to large
radii.  Hernquist \& Quinn (1987) use the remarkably regular shell system
NGC 3923 to argue that MOND predicts the wrong number of shells for the
periods implied by the observed relative radii, and also that the slope of
the radius--shell number relation is incorrect (their Fig.~2).  The latter
argument is based on normalization to the outermost shell ($N_{shell} =1$).
This is a little unfair, as only the shape is predicted.  The predicted shape
is approximately correct for $1 < N_{shell} < 17$ --- it is the outer
shell which is deviant.  Unlike the majority of the shells in this system,
this outermost shell is substantially offset from the major axis.
Alignment with the major axis is central to the argument,
which assumes the shell system was created by a simple radial phase-wrapping
merger event (Quinn 1984).
Any deviations from this simple idealized case complicates the interpretation.

The periodicity argument Hernquist \& Quinn (1987) make is more persuasive.
For a phase wrapped system, MOND should have formed more shells than they
count.  This requires that phase wrapping is an adequate model and that
nearly all shells have been detected.  Hernquist \& Quinn (1987)
argue that it would be difficult to miss the additional
shells that seem to be required by their MOND argument.
However, Prieur (1988) did discover additional shells
which had previously been missed.

Though a good argument, the period-number relation of Hernquist \& Quinn
(1987) does not constitute the direct test of the MOND force law.
The argument is based on a very
idealized realization of simple phase-wrapping as the result of a minor
merger.  Prieur (1988) shows that NGC 3923 is not as clean a system
as required by this picture.  Indeed, Hernquist \& Spergel (1992)
suggest that such rippled systems can result from {\it major\/} mergers
(as already suggested from observational evidence by Schweizer 1982 and
McGaugh \& Bothun 1990).  They specifically cite NGC 3923 as a likely example.
Even in the conventional context, they abandon the minor merger phase wrapping
hypothesis which is the basis of the argument against MOND.

Another interesting argument involving an elliptical galaxy is the apparent
difference between the shape and orientation of the optical and X-ray
isophotes in NGC 720 (Buote \& Canizares 1994).  These authors point out
that in any modified theory of gravity, the isopotential surface presumably
traced by the gas should not differ from that determined by the dominant
stars.  Any significant difference would impose a geometrical requirement for
a dark mass component.

In other elliptical galaxies Buote \& Canizares (1994) examine there is
no apparent difference in this sense.  However, in the case of NGC 720,
the stars have an ellipticity
$\epsilon_* = 0.4$ and the gas $\epsilon_g = 0.2$ -- 0.3 with a difference
in position angle of $30 \pm 15^{\circ}$ (Buote \& Canizares 1994, 1996).
The X-ray isophotes twist and appear to become more pointy than the optical
isophotes outside $\sim 1 R_e$.  Buote \& Canizares (1994) argue
that the potential due to the stars can only become rounder with increasing
radius, so the pointy X-ray isophotes provide geometrical evidence in favor
of dark matter.

This situation is rather puzzling even with dark matter, since
the stars should still contribute substantially to the mass over the observed
region.  Even though the X-ray isophotes appear to become more elongated
than those of the stars, Buote \& Canizares (1997) derive a dark matter
mass potential which is more round than either stars or gas.  This is less
easy to accomplish in MOND since there is no freedom to vary the shape and
position angle of the dominant mass as there is with dark matter.
Nevertheless, both dark matter and MOND imply potentials consistent with
or slightly rounder than the isophotes of the stars and rounder than the
X-ray isophotes.  The interpretation of NGC 720 seems difficult with either
dark matter or MOND if we accept the data at face value.

The basic argument of Buote \& Canizares (1994) is based on the difference
between the X-ray and optical isophotes.  The X-ray isophotes are very ragged
(see Fig.~2 of Buote \& Canizares 1994 or Fig.~1 of Buote \& Canizares 1996)
and do not obviously provide a strong constraint.  Paper III provides several
examples of how misleading the shapes of ragged isophotes can be.

The geometrical argument of Buote \& Canizares (1994) is valid in principle.
Observing many more galaxies with considerably higher signal to noise
would prove interesting.  At the present time,
these and other data for giant elliptical galaxies provide only weak and
indirect tests of MOND.  None of these data clearly contradict it.

\subsection{Disk Stability}

Another indication of a mass discrepancy is in the long term existence
of dynamically cold spiral disks.  Purely Newtonian disks are subject to
global instabilities which rapidly lead to their demise unless
stabilized by a dominant dark halo (Ostriker \& Peebles 1973).  What really
matters here is the ratio of binding to kinetic energy; this can be
achieved either with dark matter of by altering the force-law.
Milgrom (1989) showed analytically that MOND disks are somewhat more
stable than purely Newtonian ones.

It turns out to be very difficult to adopt standard N-body codes to address
this problem properly (Mihos, private communication).  Brada (1996) developed
an alternative approach based on the multigrid algorithm.  This supports and
extends the analytic conclusions reached by Milgrom (1989):  MOND disks
are more stable than purely Newtonian disks both locally and globally.
The additional stability is fairly modest, roughly equivalent to that
provided by a halo with $\mass(R=5h) \approx 3 \mass_{disk}$ (Brada 1996).

This leads to a further test.  The stability properties predicted by dark
matter and MOND diverge as surface brightness decreases.  In the case of dark
matter, the halo mass enclosed by the disk increases systematically with
decreasing surface brightness (Fig.~4b of paper I).  Self-gravity in the
dynamically cold disk
is the driving force for bars and spiral arms, but becomes progressively less
important.  At some point such features should be completely suppressed
by the dominant hot halo.  In contrast, the amount of additional stability
provided by MOND depends only weakly on surface density and the self-gravity
of the disk always matters.

Spiral features appear feeble in LSB galaxies, but are clearly
present (Schombert \etal 1992; McGaugh \etal 1995;
de Blok \etal 1995; Impey \etal 1997).  Brada
(1996) predicts the growth rate of the $m = 2$ mode for both dark matter and
MOND (his Fig.~11).  Using the observed accelerations as a scale, LSB galaxies
correspond to his mass models with $\mass < 0.1$  Below this level,
instabilities are almost completely suppressed in the dark matter case
(see also Mihos \etal 1997).

In order to generate spiral structure internally, the disk needs to be
rather heavy (Athanassoula \etal 1987).  In LSB disks, it is conceivable
that the minimum disk mass required to generate spiral arms might
exceed the maximum disk mass allowed by the rotation curve (see
Quillen \& Pickering 1997).  This may provide a further test.

It may be that spiral arms do not have an origin internal to the dynamics
of the disk.  However, it is difficult to invoke an external trigger
in LSB galaxies since they are quite isolated systems (Bothun \etal 1993;
Mo \etal 1994).  If spiral structure has nothing to do with disk kinematics,
then of course there is no test one way or the other.

The form of the surface brightness distribution in the field (McGaugh 1996;
Fig.~8 of paper I)
may also hold a clue.  Milgrom (1989) argued that the stability of disks
would lead to a transition at the critical surface density
$\sigma_0^* \approx a_0 G^{-1} = 880 \mass\solar {\rm pc}^{-2}$.  Above this,
the typical acceleration of a disk will exceed $a_0$ and one expects purely
Newtonian behavior.  Since bare Newtonian disks are subject to self destructive
instabilities, they should not survive.  Below the critical surface
density, MOND lends the extra stability discussed above and a disk may
assume any $\sigma_0 < \sigma_0^*$.

There should therefore exist a sharp cut off in the surface brightness
distribution at $\lincsb^* = \sigma_0^*/\ML*$.  Such a sharp cut off is
observed, though its precise position is rather uncertain (McGaugh 1996).
For $\ML* = 2$, one expects $\csb^* = 20.4$, consistent with the bright
edge of the surface brightness distribution determined by de Jong (1996).

The distinction between giant ellipticals and spirals may be dictated by the
critical $\sigma_0^*$.  Gas dissipation tends naturally to form
disk systems.  However, these are only stable if $\sigma_0 < \sigma_0^*$.
An elliptical galaxy may result when, whether through initial
conditions or subsequent mergers, a system exceeds the critical surface
density.  It will not be stable as a disk, but since there is no objection to
three dimensional, pressure supported, Newtonian systems, an elliptical
galaxy seems a natural result.  The collapse of rotational support presumably
has drastic consequences for the gaseous content of such a system, so one
might expect a starburst to consume the gas roughly coeval with the
instability event.  This would leave the signature of a
population dating to said event, so that at present elliptical galaxies
would appear gas poor and old.

This same process could also occur with conventional dynamics and
dark matter halos.  Should self-gravity become sufficiently great in high
surface density disks, the halos will no longer suffice to stabilize them.
However, the dark matter picture offers no reason why this should happen
at the particular scale $a_0 G^{-1}$ natural to MOND.

The thickness and velocity dispersion of disks may provide a further test.
In the Newtonian case the dark mass is arranged in a halo,
so the thickness of the disk is determined solely by the mass of the stars and
the usual conventional dynamical equations.  For a vertical density distribution
\begin{equation}
\rho(z) = \rho(0) {\rm sech}^2\left(\frac{z}{z_0}\right)
\end{equation}
(Spitzer 1942),
\begin{equation}
z_0 = \frac{\varsigma_z^2}{\pi G \sigma}.
\end{equation}
This would lead one to expect the
disks of LSB galaxies either to be relatively thick or to have very low
vertical velocity dispersions since the low surface mass
density disks have little self-gravity.  In contrast, MOND increases the
binding force over the Newtonian prediction in a way which increases
with decreasing surface density.  The modified version of the expression
for disk thickness (Milgrom 1983b) becomes
\begin{equation}
z_0 = \mu(x) \frac{\varsigma_z^2}{\pi G \sigma}
\end{equation}
where $\mu(x)$ is the interpolation function of MOND.
Disks in the far MOND regime should be thinner than the Newtonian
equivalent by the factor $\mu(x) \approx x = a/a_0$ for a given velocity
dispersion.  For a given disk thickness, MOND disks can support a
velocity dispersion a factor of $\sqrt{\mu(x)}$ higher.

This could provide a strong test.  Low surface brightness disks appear to
be quite thin (Dalcanton \& Schectman 1996; Kudrya \etal 1994).  If
they are Newtonian they must have quite small velocity dispersions.
MOND disks of the same thickness would have distinctly higher $\varsigma_z$.

Moreover, there comes a surface density where Newtonian disks cease to be
disks at all.  To illustrate this, consider the disk thickness resulting
from equations~(27) and (28) with velocity dispersions plausible for
the central regions of low surface brightness disks.  The only difference
between these two equations is the MOND interpolation function $\mu(x)$.
For plotting convenience we adopt $\mu(x) = x/\sqrt{1+x^2}$ (Milgrom 1983a).
The interesting effects occur in the asymptotic regime $\mu(x) \approx x$
where the assumed form of $\mu(x)$ is irrelevant.  Assuming the
central regions of the disk can be approximated as a plane parallel slab
where $V^2/R \ll \varsigma_z^2/z_0$ so that the vertical restoring force
dominates the acceleration allows us to approximate $x$ as
$x \approx \sqrt{\sigma_0/\sigma_0^*}$.  This is useful for illustrating
the dependence of the disk thickness on the central surface brightnesses
and velocity dispersion without stipulating a specific form of $V(R)$.
In general, the precise MOND prediction depends on the total acceleration,
and radial, tangential, and vertical components can all contribute.
The approximation made here using only the vertical component is adequate
for illustrating the relevant effects near the centers of disks.
The most interesting effect is the behavior behavior of purely Newtonian
disks of very low surface mass density.

The disk thickness is plotted as a function of disk central mass surface
density in Fig.~9 for plausible values of the velocity dispersion and
disk scale length.  Note that as the central mass surface density decreases,
there comes a point (which also depends on
$\varsigma_z$) where Newtonian disks rapidly become intolerably thick.
Unless $\varsigma_z$ is quite low for very LSB disks, these objects should
not be disks at all.  Disks remain reasonably thin
in MOND because the restoring force of the disk is larger.

\placefigure{schz}

Consider the actual numbers for the case illustrated in Fig.~9.
A normal sized ($h = 3$ kpc) disk
with a central velocity dispersion $\varsigma_z = 20\kms$
is respectably thin in both Newtonian and MOND cases for
$\sigma_0 > 80\,\mass\solar {\rm pc}^{-2}$ ($\csb \approx 23$ for $\ML* = 2$).
Below this surface density, the two predictions diverge.  MOND disks remain
credibly thin ($z_0/h < 1/4$) down to
$\sigma_0 \approx 2\,\mass\solar {\rm pc}^{-2}$ ($\csb \approx 27$).
In the Newtonian case, $z_0/h = 1/4$ occurs at $\sigma_0 \approx
30\,\mass\solar {\rm pc}^{-2}$ ($\csb \approx 24$) and by $\sigma_0 =
10\,\mass\solar {\rm pc}^{-2}$ ($\csb \approx 25$) the object ceases
to be a disk at all, with $z_0 \approx h$.  In order to have such low
surface brightness disks (which do exist) the central velocity dispersion
must be very low: $\varsigma_z < 10\kms$ is required to postpone the
Newtonian divergence to the regime $\csb > 25$ not yet observed.
Global stability is not the only problem for purely Newtonian disks:
it is also difficult to explain the existence of cold, thin disks of
low surface brightness with a purely Newtonian force law.

Unfortunately, no stellar velocity dispersion data exist for LSB galaxies.
This would be an extremely difficult observation, but would provide a
powerful test.  Velocity dispersion data do reach to large radii in some high
surface brightness disks (Olling 1996a).  If interpreted
in terms of dark matter, the MOND signature in such data will be a
requirement either for a large amount of disk dark matter or a
flattened halo.  Data which reach far enough to imply a large mass discrepancy
should make it necessary to put a lot of dark mass in a distribution close
to that of the disk (\eg Olling 1996b).  In very LSB disks, there may
come a point where the dark matter required to bring this about would exceed
that allowed by the maximum disk solution.

A related observation
is the Oort discrepancy in the Milky Way.  Kuijken \& Gilmore (1989)
test MOND in this context, and find that it tends to over-correct somewhat.
However, they used a value of $a_0$ nearly four times larger than currently
measured.  This will cause a mass discrepancy to be implied by the MOND
equations before it actually should be, leading to an apparent over-correction.

\subsection{Clusters of Galaxies}

Perhaps the strongest observational argument against MOND at the time of
its introduction was that while it worked in galaxies, it failed in
clusters of galaxies.  The apparent mass discrepancies $\mass_N/\mass_* \gtrsim
100$ were too large to be explained by the typical accelerations,
$a_0/\langle a \rangle \sim 10$.  Even with MOND,
substantial amounts of dark matter seemed to be required, an unacceptable
situation.

The basic picture which held for many years is that the amount of dark
matter increased with increasing scale.  Mass discrepancies in
galaxies were a factor of ten, those in clusters factors of hundreds,
and even more was required to close the universe.  This picture
of ever more dark matter on ever larger scales has changed
dramatically in recent years.  X-ray observations of rich clusters of
galaxies have shown that much or even most of the baryonic matter is
in the form of hot gas (\eg David \etal 1990; White \& Fabian 1995).
Rather than having more dark matter than individual galaxies, clusters are
in fact more baryon dominated (paper I).  There were in effect {\it two\/}
missing mass problems in clusters: the usual dynamical mass discrepancy and
the fact that many of the baryons are in the form of hot gas rather than
stars.

The need for an additional mass component which
seemed so unacceptable for MOND would now appear
to be a successful prediction.  It is thus important to reanalyze
MOND in light of this new knowledge.  The \& White (1988) did this for
the case of the Coma cluster, and found that MOND still seemed to be off
by a factor of $\sim 2$.  Recall, however, the many
uncertainties associated with tests in quasi-spherical systems discussed in the
case of dwarf Spheroidals.  Given the uncertainty in the underlying
assumptions of sphericity and virial equilibrium upon which the analysis is
based, is a factor of two a problem or a success?

Andernach \& Tago (1997) have compiled a good deal of cluster redshift
data, and quote a median cluster velocity dispersion of 695\kms.
For a spherical virialized system, this corresponds to a MOND mass
of $3 \times 10^{14} \mass\solar$ (equation~22).  The typical hot gas
mass of clusters in the compilation of White \& Fabian (1995) is
$\sim 10^{14} \mass\solar$.  So even a crude calculation comes pretty close,
especially if there is a comparable amount of mass in stars.

Sanders (1994) re-addressed the problem in greater detail.
For 16 clusters, he shows that the MOND mass is strongly correlated
with the X-ray gas mass.  The MOND masses are a bit greater than the
gas mass (his Fig.~3) indicating some additional mass in stars.

A more thorough investigation of this particular test would require a great
deal more data than are available to us:  combined X-ray and optical
observations giving good estimates of the mass in each component
and accurate velocity dispersions.  Obtaining the full optical luminosity
of a cluster is far from trivial (Impey \etal 1988), and X-ray observations
still lack the spatial resolution to guarantee the validity of the assumption
of isothermality.  Obtaining a reliable velocity dispersion from a sample
of galaxies that are truly virialized is also challenging.  Interlopers
and substructure could play a strong role is distorting the mass-indicative
velocity dispersion.  A 40\% overestimate of $\varsigma$ would imply a
MOND mass a factor of 4 too great.  This would appear disastrous if this
uncertainty is not considered and properly propagated.
It is important to test MOND in clusters as adequate data become available,
keeping in mind the many uncertainties.

\subsection{Gravitational Lensing}

Another indication of a mass discrepancy independent of kinematic data
is gravitational lensing, both in individual galaxies (\eg Kochanek 1995)
and clusters (\eg Tyson \etal 1990).  Since there is not yet a relativistic
extension of MOND, there is no clear prediction for gravitational lensing.
That lensing occurs must ultimately be explained, but at present it provides
no test of the validity of the MOND force law.

Some progress has nevertheless been made.  If MOND is interpreted as an
alteration of the law of inertia, it is fairly successful at explaining
weak lensing in clusters.  However, it does then predict that additional,
as yet undetected baryonic mass resides in the cores of a few of the most
X-ray rich clusters (Milgrom 1996).  This is apparently the only place where
MOND does not remedy the mass discrepancy problem in the sense that significant
additional mass remains hidden.  This is a very limited
missing mass problem, restricted to the central Newtonian cores of some rich
clusters.  These are generally cooling flow clusters, so at least some
additional baryonic mass is already inferred to reside there.

Gravitational lensing observations have allowed entire classes of theories
to be ruled out as the possible basis of combining relativity and MOND
(Bekenstein \& Sanders 1994).  This does not mean none can exist.
One attempt at such a theory (Sanders 1997) has an interesting consequence.
In this case, the mass distribution that would be inferred when interpreted
in terms of dark matter is the same for both lensing and kinematical
observations.  Hence, observational agreement between such observations
does not uniquely require a dark matter interpretation.

\subsection{Large Scale Structure}

Observations of the motions of galaxies now extend over very large scales.
These do not provide strong tests of MOND, since we must make
crude assumptions about the mass distribution.
We can, however, address the consistency of MOND with current observations
on a qualitative level.

That there are large ($\gtrsim 30$ Mpc) voids and filaments in the large
scale galaxy distribution is now regularly reproduced by simulations of
structure formation.  It is worth recalling that these features
initially came as a great surprise (de Lapparent \etal 1986):
the universe was suppose to be homogeneous on these large scales.
That these large, sharp features exist makes a reasonable amount of sense in
the presence of a force law which is effectively enhanced on large scales,
as does the occasional inference of an excessive mass concentration
like the Great Attractor (Lynden-Bell \etal 1988).

Where dynamical measurements exist, these large scale structures can be used as
crude tests of MOND.  Milgrom (1997) presents an analysis of the Perseus-Pisces
filament, and derives $\Upsilon \sim 10$.  This is fairly reasonable
considering the crude assumptions that must be made (\ie that the
structure is a virialized linear feature) and that it may well contain
substantial amounts of gas.

The expansion of voids could provide another test.  As underdense
regions, voids {\it must\/} expand in the conventional picture.  This need
not be the case in MOND, for which the effective potential is logarithmic
on large scales.
A spherical shell mass distribution will eventually turn around and collapse
regardless of its interior density
(Felten 1984), so one might expect to see some voids expanding and others
contracting.  In the one case where careful distance as well as redshift
measurements have been made (Bothun \etal 1992), no expansion is detected
($\lesssim 5\%$ of the void diameter in $\kms$).

On the largest scales, MOND does require a universe composed entirely of
baryons.  It may therefore seem troubling that dynamical estimates of the mass
density are persistently around $\Omega \approx 0.3$, much higher than the
baryon density allowed by primordial nucleosynthesis, $0.01 \le \Omega_b
\le 0.03$.  These estimates of $\Omega$ are of course based on
the usual Newtonian equations, and MOND will require much less mass.
As with all mass discrepancies,
the amount by which $\Omega$ is overestimated depends on the typical
acceleration scale probed:
\begin{equation}
\Omega_{\rm MOND} = \frac{\langle a \rangle}{a_0} \Omega.
\end{equation}

It is difficult to estimate $\langle a \rangle$ here, but as an example
we make the usual assumption of homogeneity so that the density field
may be approximated as a constant.
The Poisson equation with the usual $\Omega = 8 \pi G \rho/3 H_0^2$
then gives
\begin{equation}
\langle a \rangle \approx | \nabla \varphi | = \frac{1}{2}H_0^2 \Omega R,
\end{equation}
where $R$ is the scale which
the observations cover.  Davis \etal (1996) give a value $\Omega^{0.6} b =
0.6$ from data which extend reliably out to $R \approx 5000\kms$.
There should be no mass biasing in MOND (though different
populations of galaxies may be biased relative to one another)
so we assume $b = 1$.  This gives a conventional
$\Omega = 0.43$.  Taking these numbers at face value leads to
$\langle a \rangle \approx 0.026 {\rm \AA}\,{\rm s}^{-2}$ and
\begin{equation}
\Omega_{\rm MOND} \approx 0.01.
\end{equation}

Given the nature of the data and the necessary assumptions, this is
probably uncertain by at least a factor of a few.
Nevertheless, it is striking that
\begin{equation}
\Omega_{\rm MOND} \approx \Omega_b.
\end{equation}
MOND appears to adequately address the dynamical mass discrepancy problem on
even the largest scale.

\section{Discussion}

We have taken care to review previous analyses which have found fault with
MOND.  There is no evidence on any scale which clearly contradicts MOND.
Some data which are cited as contradicting MOND actually appear to support
it (\eg that for dwarf Spheroidals). It should be noted that
not all previous independent analyses of MOND have been negative.
Begeman \etal (1991), Morishima \& Saio (1995), and Sanders (1996) all
report positive tests of the theory.  The new data for LSB galaxies we have
collected obey all the predictions made by Milgrom (1983b).

The observational tests of MOND we have discussed are summarized
in Table~3.  That LSB galaxies fall on the Tully-Fisher
relation is a strong prediction of MOND (\S 2.3).  The
$\Upsilon$-$\Sigma$ conspiracy which occurs when interpreting the LSB
galaxy Tully-Fisher relation with dark matter (Zwaan \etal 1995; paper I)
can be derived from MOND (\S 2.4).
Stellar mass-to-light ratios which agree well with what is
expected for stars can be computed directly from the observations with
MOND (\S 2.5).  The mass surface density implied by the kinematic
measure $V_c^2/h$ is strongly correlated with the disk surface brightness
\csb\ with the slope expected in MOND (\S 2.6).  The radial variation of
the dynamical mass-to-light ratio computed conventionally behaves in a
manner consistent with the predictions of Milgrom (1983b).  Similarly,
the radii of transition to apparent dark matter domination and the
typical accelerations observed in disks vary with
surface brightness as expected in MOND (\S 2.7).
The systematic dependence of the shape of rotation curves on surface
brightness is also predicted by MOND (\S 2.8).  Taken in sum, the
data are well fit by MOND (\S 2.9; paper III).
Indeed, all of the empirical facts we identified in paper I describing
the systematic properties of the rotation curves of disks as a function
of surface brightness were anticipated by Milgrom (1983b).

MOND also survives tests in systems other than disks.  Dwarf Spheroidal galaxies
are an important example.  Once thought to fail there (Gerhard \& Spergel
1992), MOND now appears to do well with improved data (\S 3.1).
On the other hand, giant elliptical galaxies provide no useful test of
MOND since the accelerations they experience do not probe the MOND regime,
at least not with current observations (\S 3.2).  The stability properties
of disks appear consistent with MOND, and the velocity dispersions of thin
LSB disks could provide a very strong test (\S 3.3).  Galaxy clusters provide
another important test. Originally seeming to require
additional dark matter in clusters (Milgrom 1983c), the detection of large
amounts of hot X-ray emitting gas in clusters generally brings these into
consistency with MOND (\S 3.4; Sanders 1994).  A definitive prediction for
gravitational lensing requires a relativistic generalization of MOND
which does not yet exist (\S 3.5).  Some types of theories can be ruled out on
this basis (Bekenstein \& Sanders 1994) while others remain possible
(Sanders 1997).  On the largest scales, MOND does as well as can be expected
given the applicability of the available data (\S 3.6).  A very low
density ($\Omega \approx 0.01$), purely baryonic universe is roughly
consistent with the dynamical data which constrain $\Omega$.

\placetable{score}

Empirically, MOND is the effective force law in disk galaxies.
It appears that this may also be the case in other systems.
The reason for this phenomenology needs to be understood.

There data allow two possible interpretations.  Either
\begin{enumerate}
\item MOND is correct, or
\item Dark matter mimics the behavior of MOND, at least in disks.
\end{enumerate}
The second possibility implies a unique and powerful coupling between
dark and luminous matter.  It is possible to write down an equation
which directly links the dark matter dominated dynamics to the detailed
distribution of the luminous matter.  This provides a new observational test
of theories of disk galaxy formation within the standard dark matter paradigm.
Since MOND always fits disk galaxy rotation curves, it must
be possible to take the luminous mass distribution
predicted for any given disk by a dark matter galaxy formation theory,
apply the MOND procedure to the luminous mass only, and
thereby obtain the correct rotation curve.  If this can not be done,
the theory has failed to produce a realistic disk.

There are no clear empirical objections to the first possibility.
Milgrom (1983b) did accurately predict numerous aspects of the kinematical
properties of LSB galaxies.  This seems unlikely to have occurred by
accident, so the possibility that MOND is correct should be considered
seriously.

\acknowledgements We are grateful to Moti Milgrom, Chris Mihos, Vera Rubin,
Bob Sanders, and the referee for close reading of this manuscript and
numerous helpful conversations.

\begin{deluxetable}{lcr}
\tablewidth{0pt}
\tablecaption{MOND \ML* \label{MONDML}}
\tablehead{
\colhead{Galaxy}  & \colhead{\ML*} & \colhead{$\xi$} }
\startdata
F563--1		&	2.6	&  0.8 \nl
F563--V2        &       1.4     &  1.6  \nl
F568--1         &       1.2     &  0.7  \nl
F568--3         &       1.8     &  1.0  \nl
F568--V1        &       3.7     &  1.3  \nl
F571--V1	&    $-$0.2\tablenotemark{a} & 0.4 \nl
F574--1         &       0.01\tablenotemark{b} & 0.6  \nl
F583--1         &       0.9     &  1.3 \nl
F583--4         &       0.2     &  0.5  \nl
UGC~~128        &       1.0     &  0.5  \nl
UGC 6614        &       2.7     &  0.7  \nl
DDO~~154	&	$\sim 0$\tablenotemark{c}	&  1.2 \nl
DDO~~168        &       0.8     &  0.9  \nl
NGC~~~55        &       0.3     &  1.3  \nl
NGC~~247        &       2.2     &  1.2  \nl
NGC~~300        &       1.6     &  1.2  \nl
NGC~~801        &       1.2     &  1.1  \nl
NGC 1560        &       2.0     &  1.3  \nl
NGC 2403        &       1.5     &  2.4  \nl
NGC 2841	&	$\sim 7$\tablenotemark{c}	&  6.1 \nl
NGC 2903        &       4.8     &  5.5  \nl
NGC 2998        &       0.8     &  2.3  \nl
NGC 3109        &       0.4     &  0.8  \nl
NGC 3198        &       2.5     &  2.6  \nl
NGC 5033        &       5.4     &  2.3  \nl
NGC 5533        &       4.1     &  1.8  \nl
NGC 5585        &       0.9     &  1.6  \nl
NGC 6503        &       1.7     &  2.3  \nl
NGC 6674        &       2.9     &  2.3  \nl
NGC 7331        &       2.6     &  3.5  \nl
UGC 2259        &       2.7     &  1.7  \nl
UGC 2885        &       1.6     &  1.8 \nl
\enddata
\tablenotetext{a}{$\pm 2.5$}
%\tablenotetest{b}{becomes positive if $i = 35 \rightarrow 34$}
\tablenotetext{b}{$\pm 1.4$}
\tablenotetext{c}{distance sensitive}
%MOND masses reduced by a factor of 1.31 to account for disk geometry at R = 4h.
\end{deluxetable}

\begin{deluxetable}{lccc}
\tablewidth{0pt}
\tablecaption{Dwarf Spheroidals \label{dwarf}}
\tablehead{
\colhead{Galaxy} & \colhead{\ML*} & {$\eta$} }
\startdata
Carina & \phn 5.4 & 0.5 \nl
Draco & 10.8 & 0.9 \nl
Fornax & \phn 0.4 & 0.6 \nl
Leo I & \phn 2.1 & 3.8 \nl
Leo II & \phn 2.5 & 1.4 \nl
Sculptor & \phn 1.2 & 0.6 \nl
Sextans & \phn 2.9 & 0.3 \nl
Ursa Minor & 16.9\tablenotemark{a} & 0.5 \nl
\enddata
\tablenotetext{a}{$\ML* = 10^{+6}_{-4}$ for an alternative measurement of
$\varsigma$.}
\end{deluxetable}

\begin{deluxetable}{lc}
\tablewidth{0pt}
\tablecaption{Tests of Predictions \label{score}}
\tablehead{
\colhead{Observational Test} & {MOND} }
\startdata
LSBG Tully-Fisher Relation & $\surd$ \nl
$\Upsilon$-$\Sigma$ Relation & $\surd$ \nl
Stellar Mass-to-Light Ratios & $\surd$ \nl
Mass Surface Densities & $\surd$ \nl
Conventional $\MLo(R)$ & $\surd$ \nl
Transition Radii & $\surd$ \nl
Characteristic Accelerations & $\surd$ \nl
Rotation Curve Shapes & $\surd$ \nl
Rotation Curve Rate of Rise & $\surd$ \nl
Rotation Curve Fits & $\surd$ \nl
Disk Stability & ? \nl
Dwarf Spheroidal Galaxies & $\surd$ \nl
Giant Elliptical Galaxies & NT \nl
Galaxy Clusters & ? \nl
Gravitational Lensing & NP \nl
Large Scale Structure & ? \nl
$\Omega = \Omega_b$? & ? \nl
\enddata
\tablecomments{\\ \centerline{$\surd$ = prediction confirmed} \\
\centerline{X = prediction falsified} \\
\centerline{? = remains uncertain} \\
\centerline{NP = no prediction} \\
\centerline{NT = no test} }
\end{deluxetable}

\clearpage

%1a,b,c
\figcaption[MLMmond.ps]{The stellar mass-to-light ratios of spiral galaxies
determined from MOND, plotted against (a) absolute magnitude, (b) central
surface brightness, and (c) scale length.  The median value is $\ML* = 1.6$.
This Figure is analogous to Fig.~4 of paper I.  As was done there,
error bars in \ML*\ have been computed assuming a nominal inclination
uncertainty of 3 degrees.  The errors are larger than those in Fig.~4 of
paper I because MOND masses are proportional to the fourth power of the
velocity, and so have a factor $\sin^{-4}(i)$ contributing to the error
instead of just $\sin^{-2}(i)$ in the conventional case.
\label{MLmond}}

%2
\figcaption[Ximond.ps]{The mass surface density $\xi$ indicated by MOND
plotted against surface brightness.  The two are strongly correlated, as
expected in the absence of dark matter.  The line is not a fit to the data.
It illustrates the slope expected from the form of the MOND force law.
The normalization is also predicted modulo the baryonic mass-to-light
ratio $\Upsilon_b$ (equation~20).  The line is drawn for $\Upsilon_b = 4$.
\label{Ximond}}

%3
\figcaption[MLR.ps]{The accumulated conventional dynamical mass-to-light ratio
($\MLo \propto V^2 R/L$) as a function of radius.  The plot is divided into
bins of different central surface brightness as indicated in each panel.
All available data (Tables 1 and 2 of paper I) have been used.
In MOND, one expects the mass discrepancy indicated by \MLo\ to be larger
and to set in at smaller radii in galaxies of lower surface brightness.
\label{MLR}}

%4
\figcaption[AR.ps]{The rotation curves plotted in terms of the requisite
centripetal acceleration $a = V^2(R)/R$.  The plot is divided into different
bins in central surface brightness, and the value of $a_0$ is marked by
the dashed line.  In MOND, one expects the acceleration
to decline with surface brightness.
\label{AR}}

%5
\figcaption[shape.ps]{The rotation curves plotted logarithmically to illustrate
their shapes.  This test is not inclination dependent.
In MOND, one expects high surface brightness galaxies to have
rapidly rising rotation curves which fall gradually to the asymptotically
flat value.  Low surface brightness galaxies should have rotation curves
which rise slowly to the asymptotically flat value.
\label{shape}}

%6
\figcaption[f5631.ps]{An example of the MOND fits to the rotation curves
of low surface brightness galaxies from paper III.  The data points are
for the galaxy F563--1.  The dashed line shows the Newtonian rotation
curve of the stellar disk and the dotted line that of the gas.  The solid
line is the resulting MOND fit to the entire rotation curve.  This
follows directly when the MOND force law is applied to
the observed luminous mass distribution.
\label{f5631}}

%7
\figcaption[aca0.ps]{The residuals (in percent) of all the MOND fits to
the rotation curves of LSB galaxies.  The difference between the observed
velocity and that predicted by MOND from the observed luminous mass
distribution ($\Delta V = V_{obs} - V_{MOND}$) is shown as a function
of the critical parameter $x = a/a_0$ where in practice $a = V^2/R$.
Each point represents one measured point from the rotation curve fits
described in paper III.  The residuals for all points from all LSB galaxies
are shown together.  This represents 15 new LSB galaxy fits with nearly 100
total independent measured points.
\label{aca0}}

%8
\figcaption[dsph.ps]{The MOND mass-to-light ratios inferred for Local Group
dwarf Spheroidal galaxies.  Triangles are the determinations of
Gerhard (1994).  Solid triangles are the values he finds for the isolated
case appropriate if $\eta > 1$ (equation~24), and open triangles are his
quasi-Newtonian values which are more appropriate when the external field is
dominant ($\eta < 1$).  Lines capped by crosses illustrate the range of
allowable values determined by Milgrom (1995). The results of Gerhard and of
Milgrom have been scaled to the value of $a_0$ adopted here which increases
their \ML*\ by a factor of 1.6.  Solid
circles are our own determinations based on the data compiled by Mateo (1996).
Dashed lines delimit the most plausible range, $1 < \ML* < 6$.
Most galaxies fall in this range according to all three independent
determinations.  In only one case (Ursa Minor) is there a marginally
significant deviation from the most plausible range.
\label{dsph}}

%9
\figcaption[SCHZ.ps]{The thickness $z_0/h$ expected for disks of various
central surface densities $\sigma_0$.  Shown along
the top axis is the equivalent $B$-band central surface brightness
\csb\ for $\ML* = 2$.  Parameters chosen for illustration are noted
in the Figure (a typical scale length $h$ and two choices of central vertical
velocity dispersion $\varsigma_z$).  Other plausible values give similar
results (equations 27 and 28).  The solid lines are the Newtonian expectation
and the dashed lines that of MOND.  The only difference between the two
cases is the MOND interpolation function $\mu(x) = x/\protect\sqrt{1+x^2}$
with $x = \protect\sqrt{\sigma_0/\sigma_0^*}$.
The Newtonian and MOND cases are similar at high surface
densities but differ enormously at low surface densities.
Newtonian disks become unacceptably thick unless LSB disks are
very cold ($\varsigma_z < 10\kms$).
In contrast, MOND disks remain reasonably thin to quite
low surface density.
\label{schz}}

\clearpage
\begin{figure}
\plotone{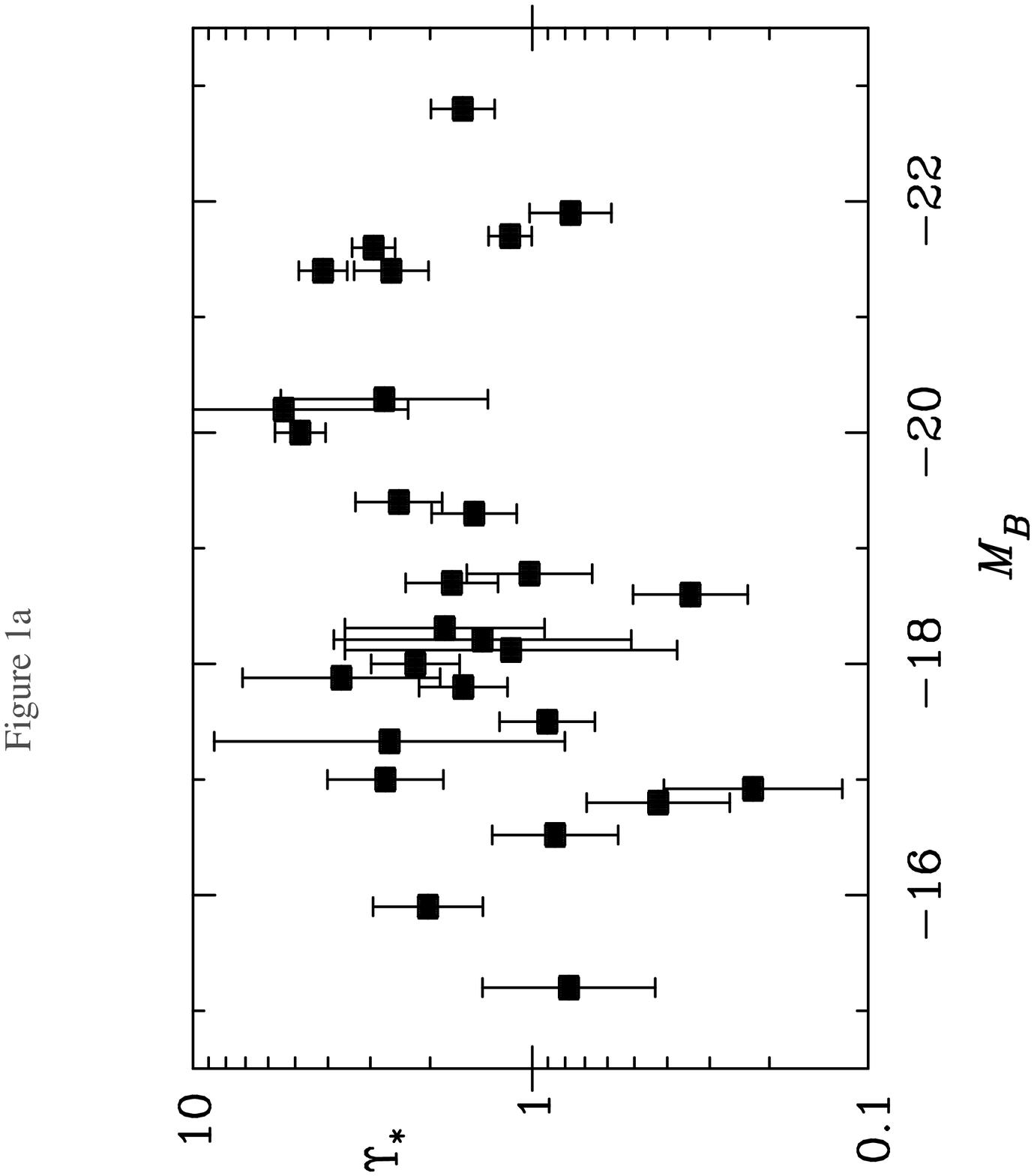}
\end{figure}

\clearpage
\begin{figure}
\plotone{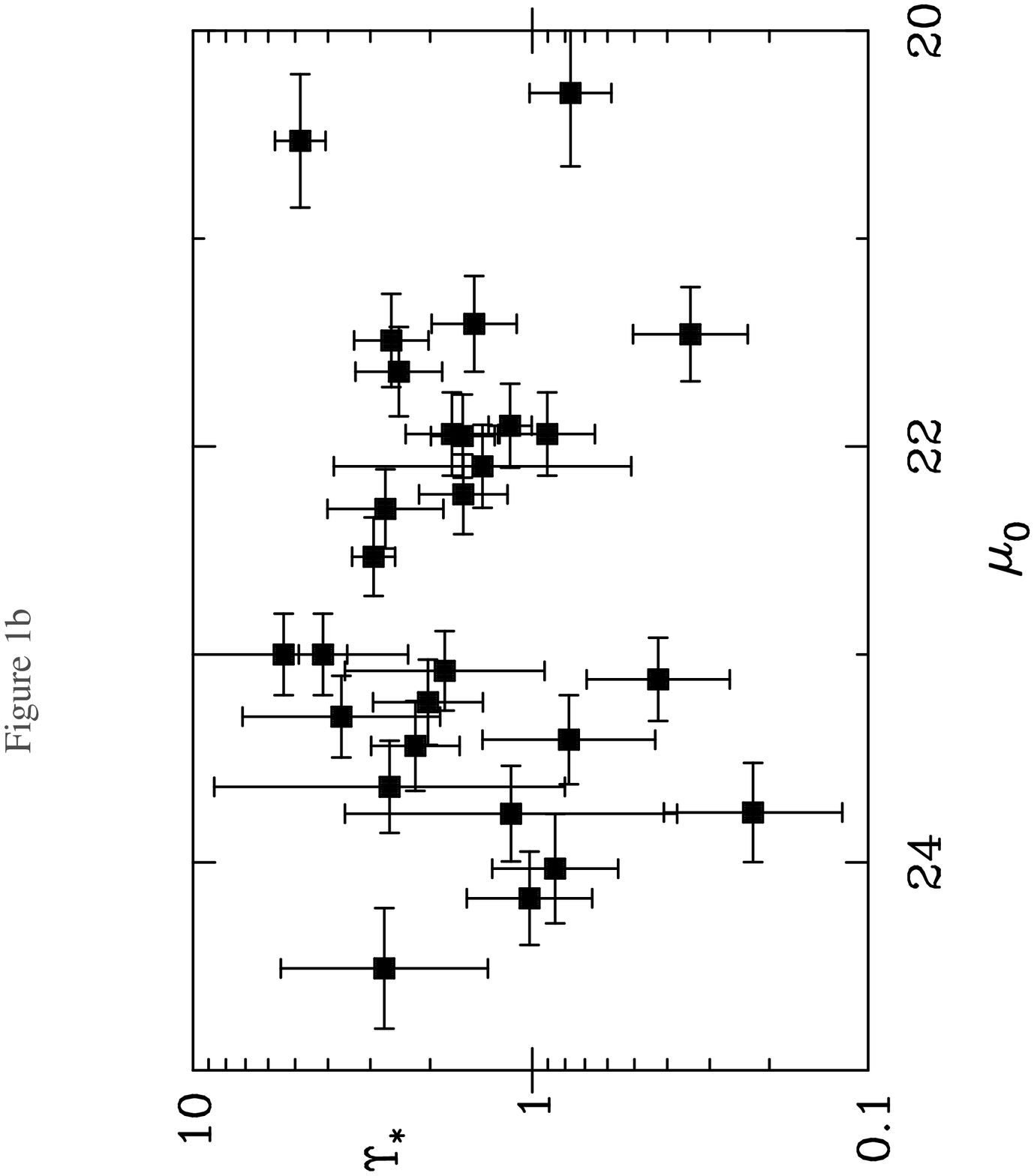}
\end{figure}

\clearpage
\begin{figure}
\plotone{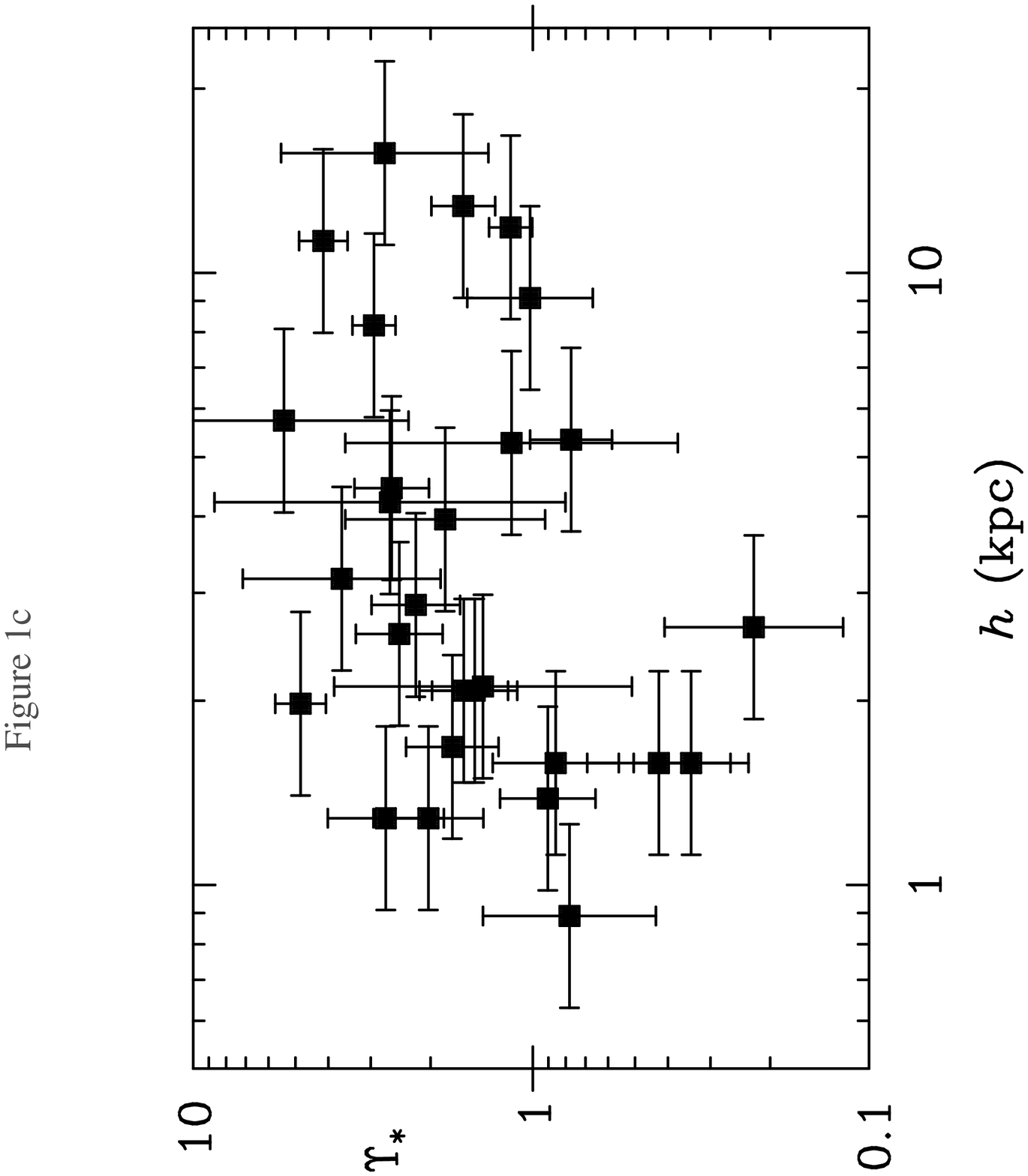}
\end{figure}

\clearpage
\begin{figure}
\plotone{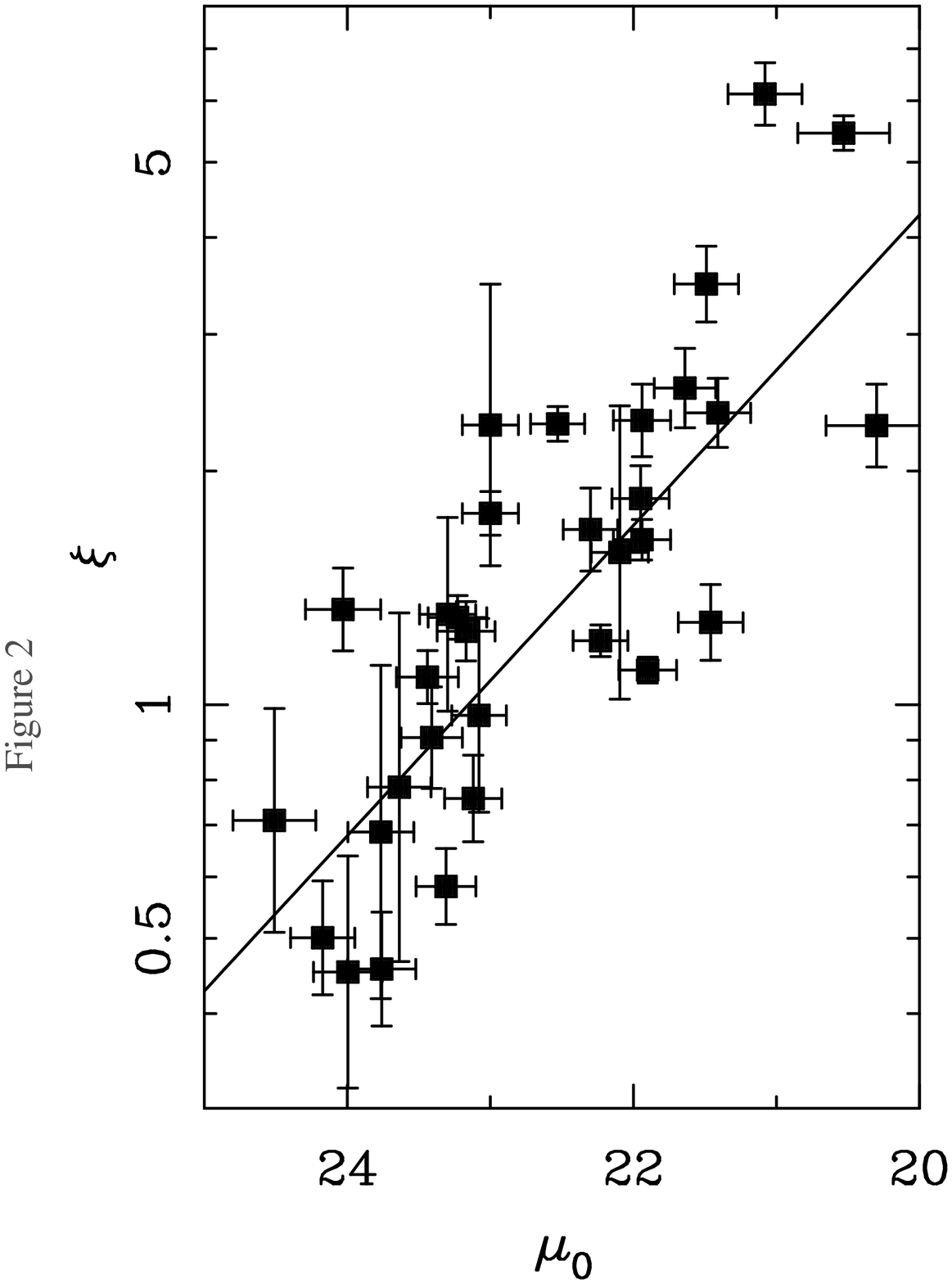}
\end{figure}

\clearpage
\begin{figure}
\plotone{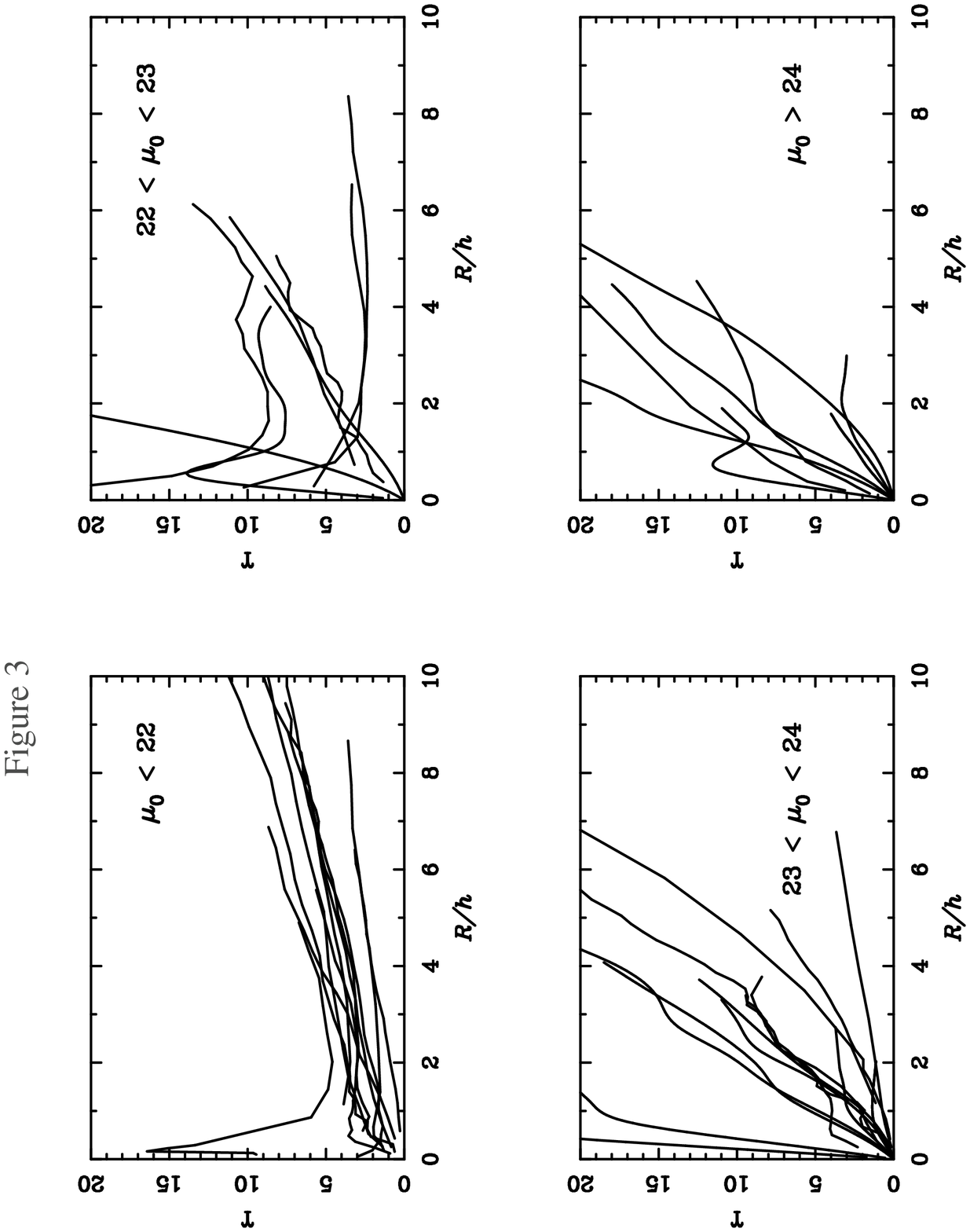}
\end{figure}

\clearpage
\begin{figure}
\plotone{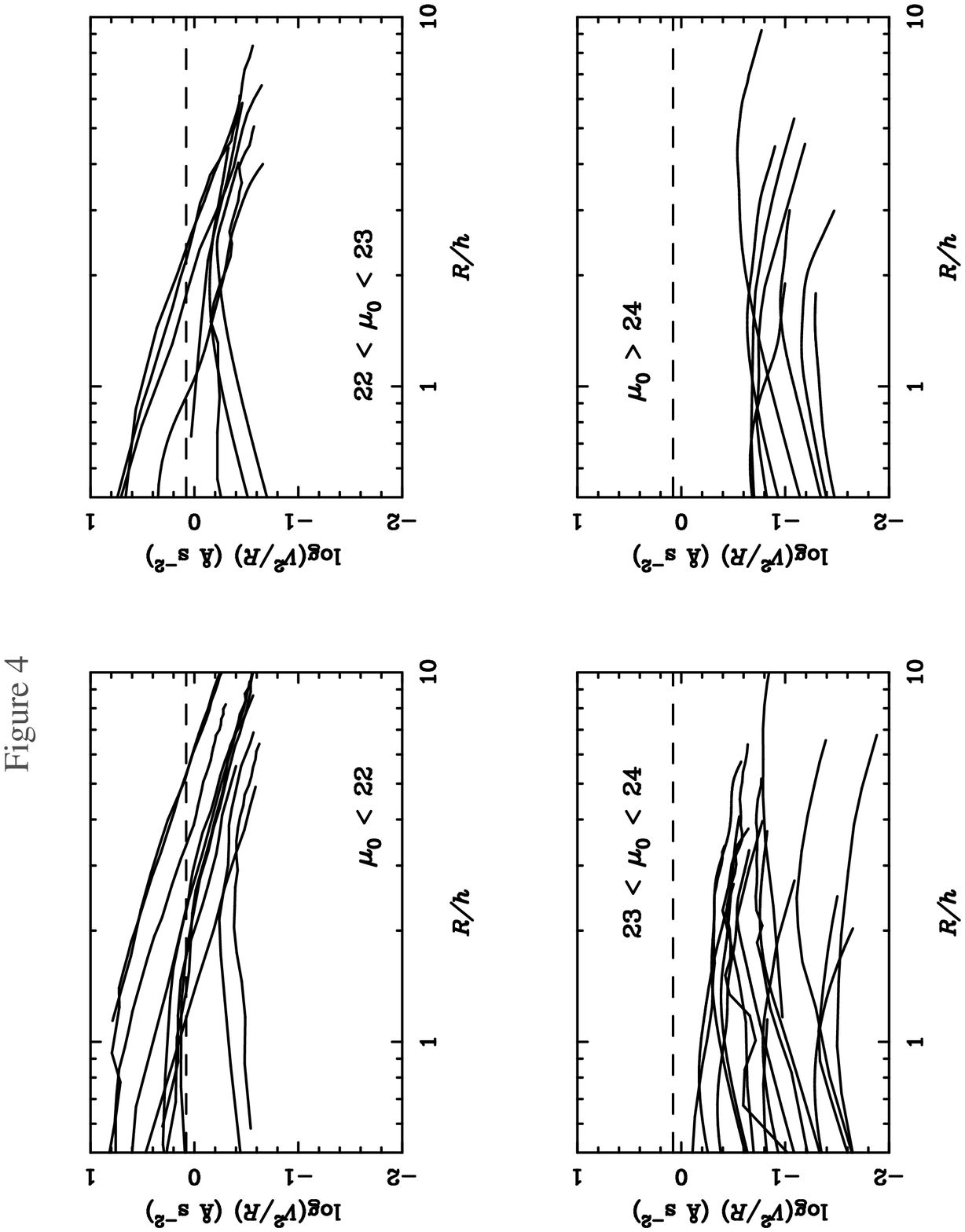}
\end{figure}

\clearpage
\begin{figure}
\plotone{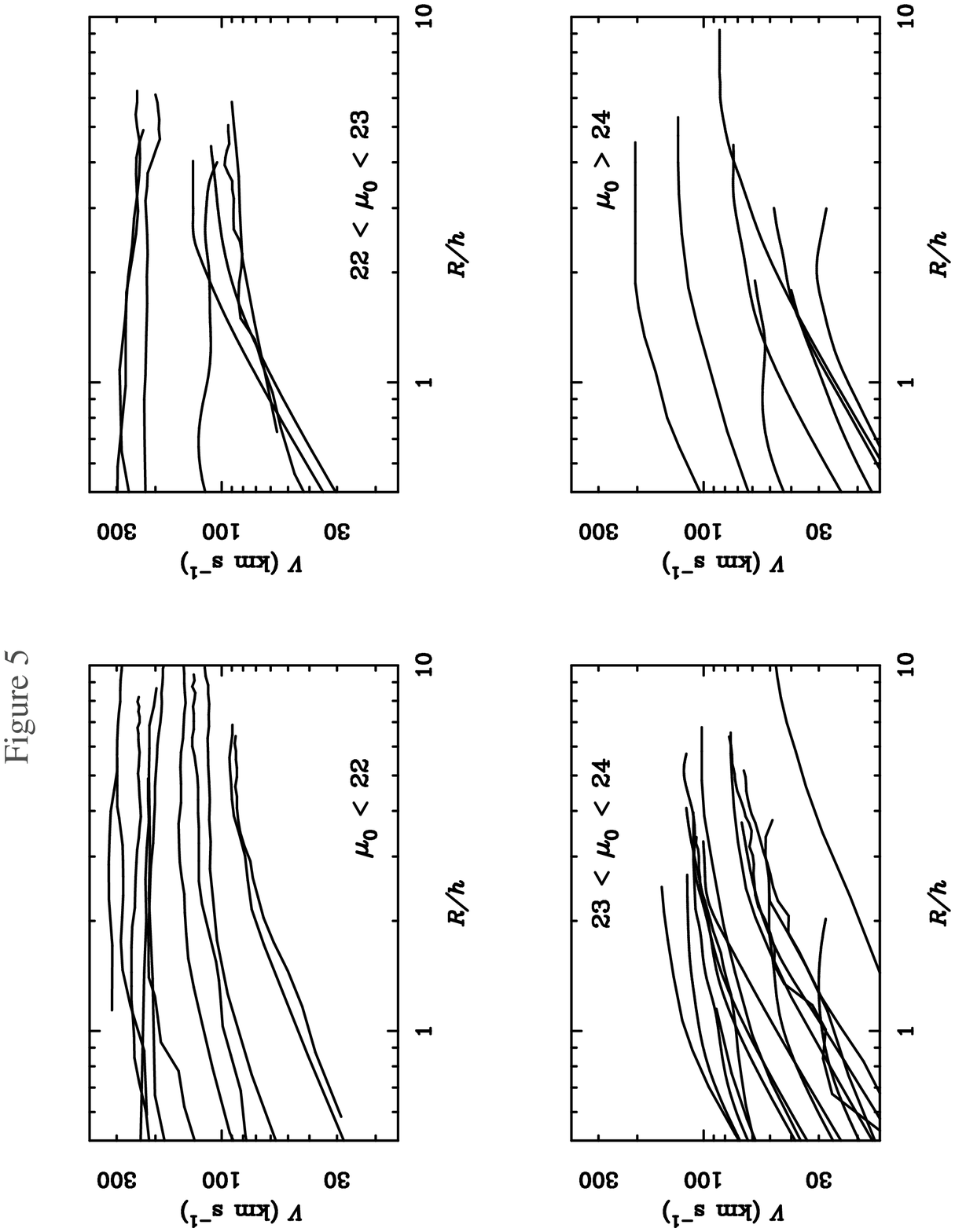}
\end{figure}

\clearpage
\begin{figure}
\plotone{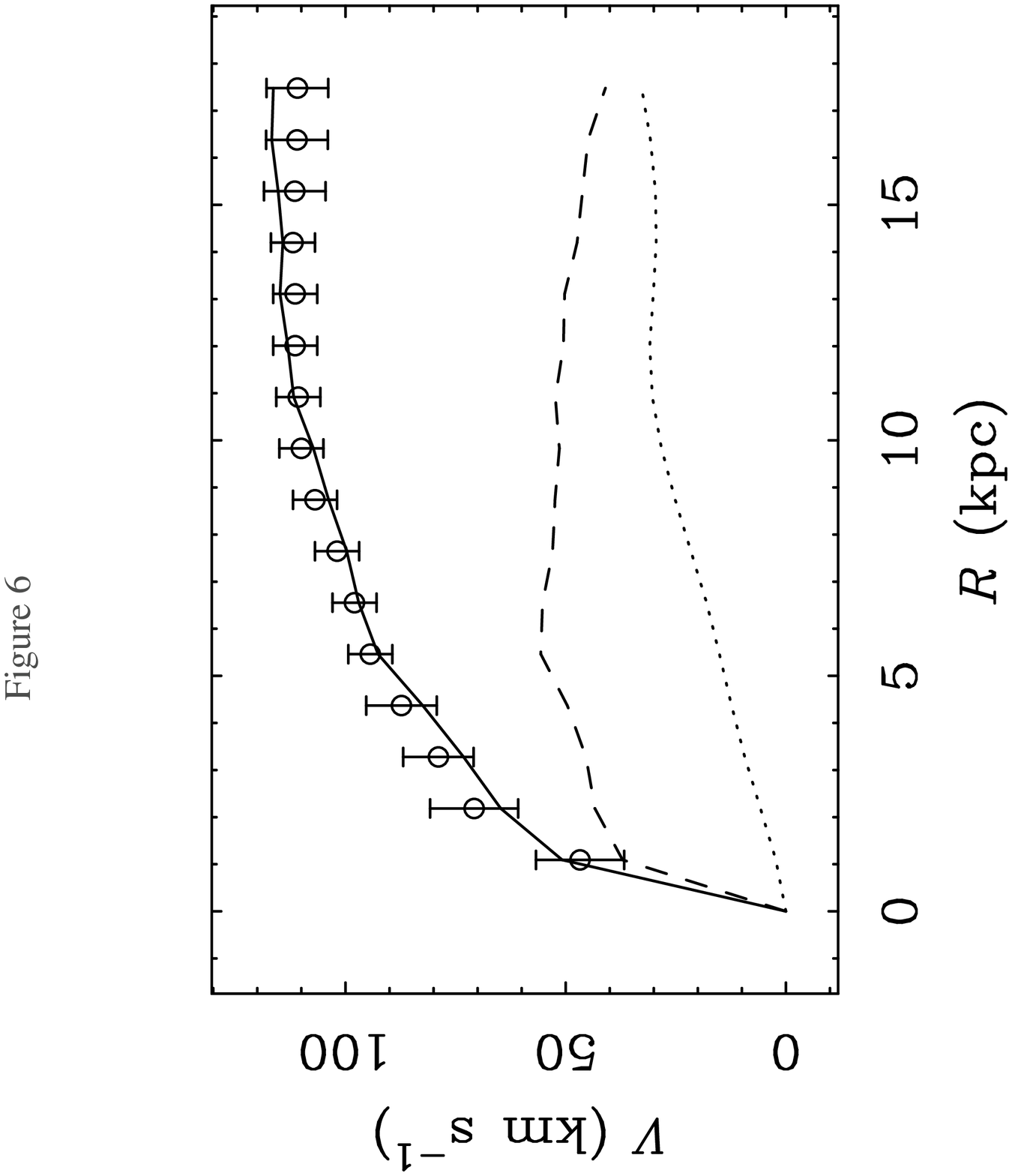}
\end{figure}

\clearpage
\begin{figure}
\plotone{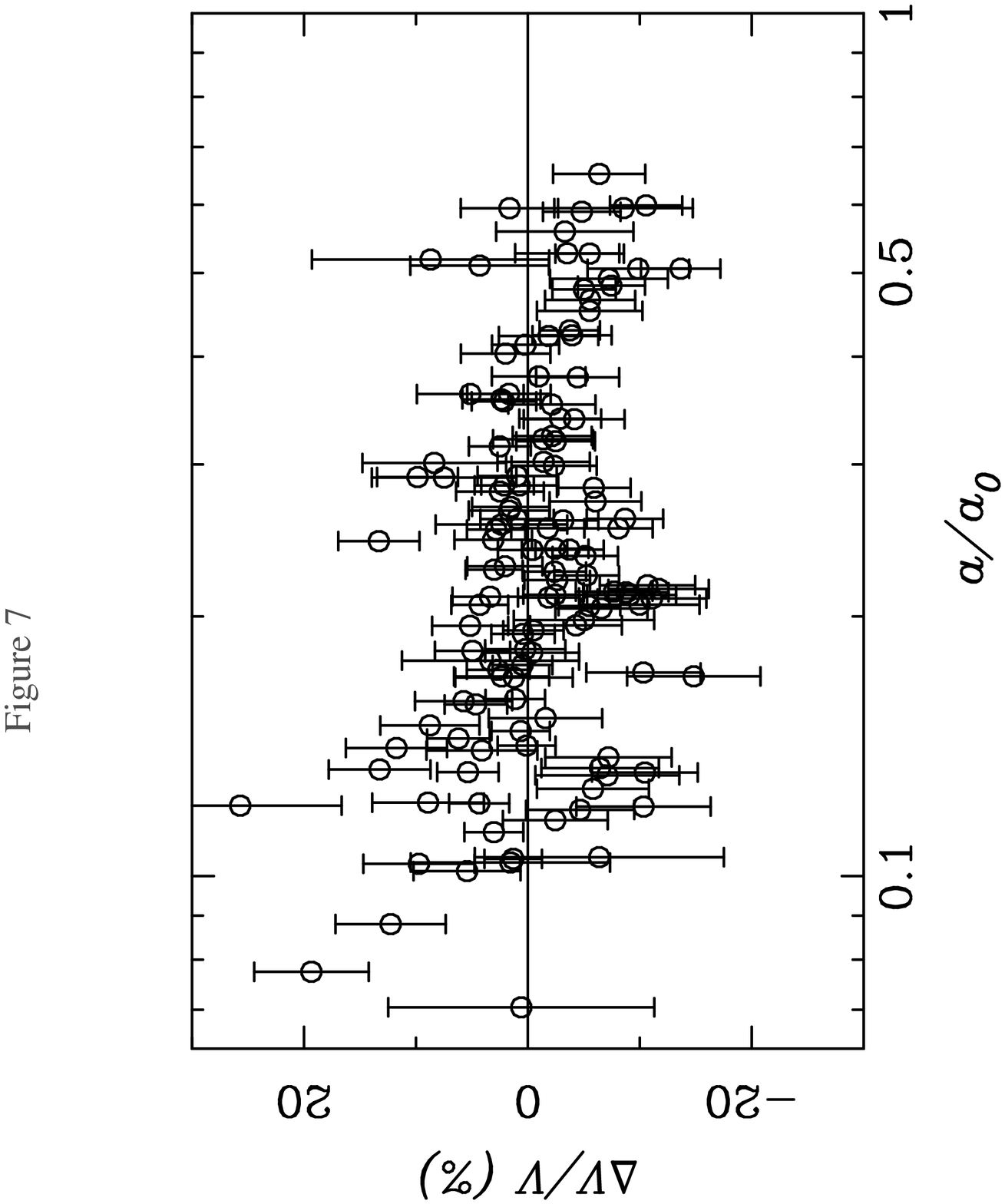}
\end{figure}

\clearpage
\begin{figure}
\plotone{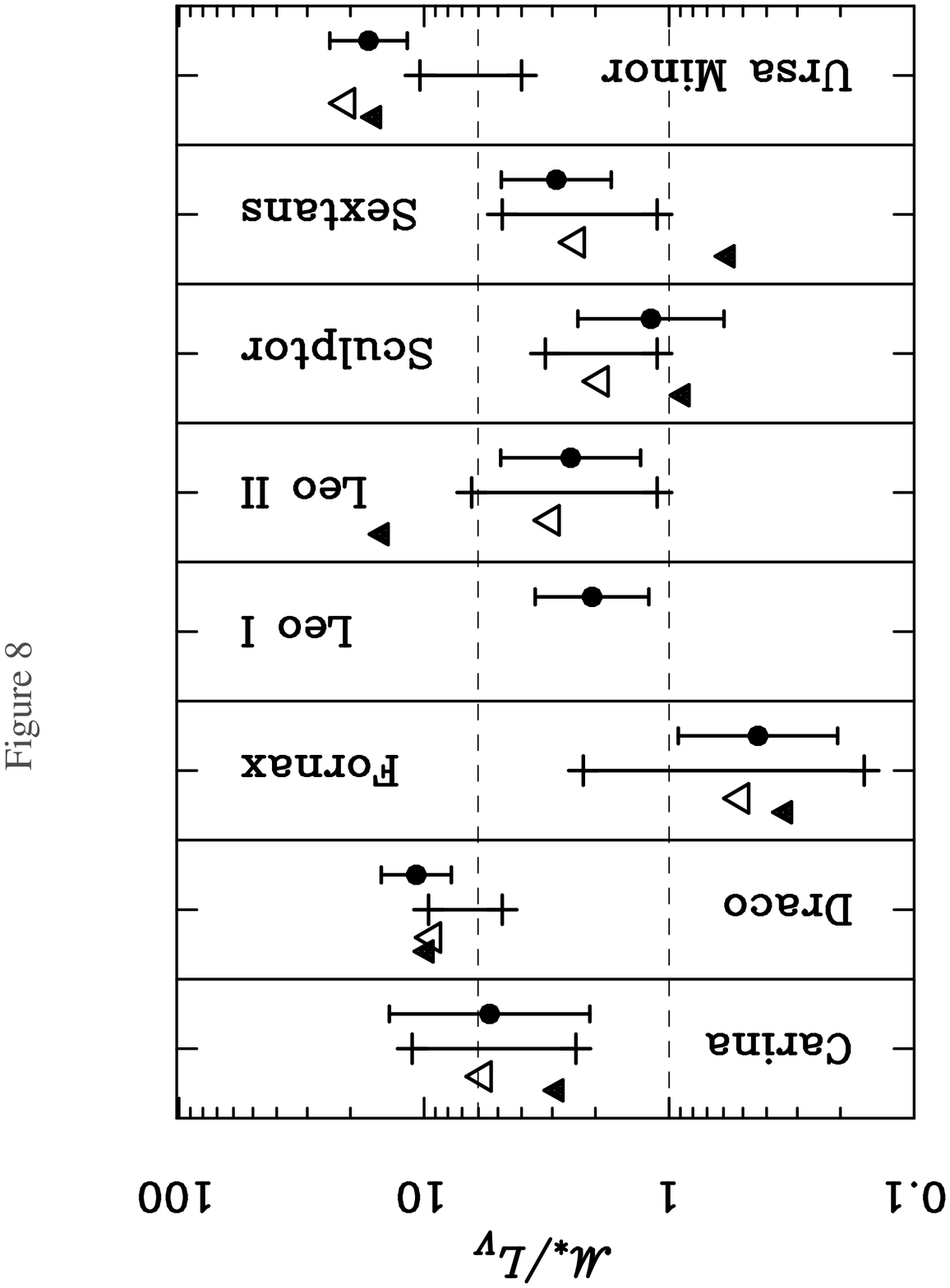}
\end{figure}

\clearpage
\begin{figure}
\plotone{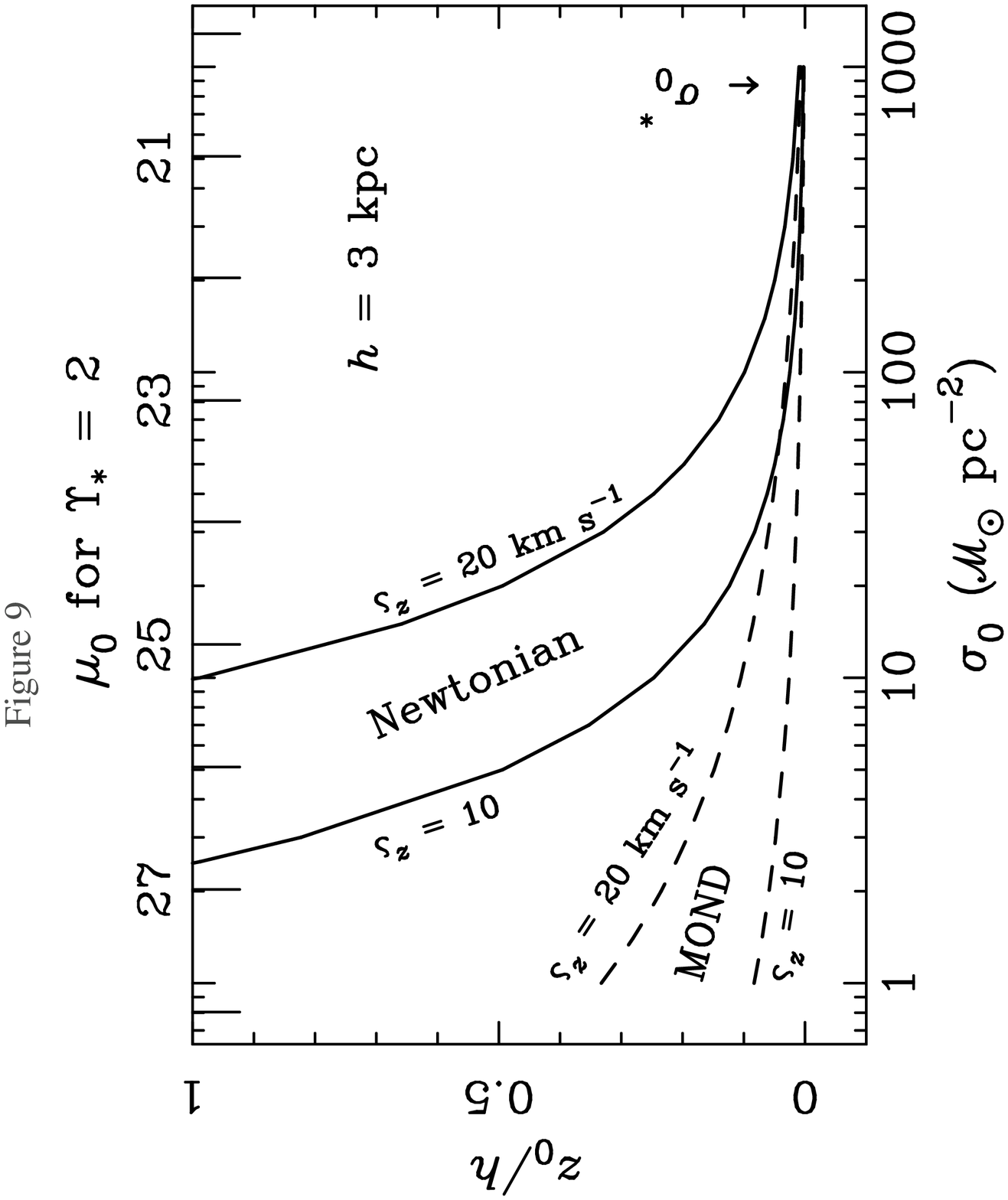}
\end{figure}

\end{document}